\documentclass[12pt]{article}
\usepackage{graphicx}
\usepackage{cite}
\usepackage{amsmath}
\usepackage{color}
\usepackage{hyperref}
\usepackage{authblk}
\usepackage{subfigure}
\usepackage{comment}
\usepackage[titletoc,title]{appendix}
\usepackage{booktabs} 
\usepackage[labelfont=bf]{caption} 

\definecolor{darkblue}{rgb}{0,0,.7}
\definecolor{darkred}{rgb}{0.7,0,0}

\hypersetup{colorlinks=true, linkcolor=darkblue,citecolor=darkblue,urlcolor=darkblue}

\numberwithin{equation}{section}

\title{Electric potential and field calculation of charged BEM triangles and  rectangles by Gaussian cubature}

\author[1,2]{Ferenc  Gl\"uck \thanks{corresponding author: ferenc.glueck@kit.edu}}
\author[1] {Daniel Hilk}
\affil[1]{Karlsruhe Institute of Technology, IKP, 76021 Karlsruhe, POB 3640, Germany}
\affil[2]{Wigner Research Center for Physics, H-1525 Budapest, POB 49, Hungary}

\date{}

\begin{document}

\maketitle

\begin{abstract}
 
It is a widely held view that analytical integration is more accurate than the numerical one.
In some special cases, however, numerical integration can be more advantageous than analytical
integration. In our paper we show this benefit for the case of electric potential and field
computation of charged triangles and rectangles applied in the boundary element method (BEM).
Analytical potential and field formulas are rather complicated (even in the simplest case of constant
charge densities), they have usually large computation times, and at field points far from the elements
 they suffer from large rounding errors.
On the other hand, Gaussian cubature, which is an efficient numerical integration method, yields
simple and fast potential and field formulas that are very accurate far from the elements. The simplicity
of the method is demonstrated by the physical picture: the triangles and rectangles with their continuous
charge distributions are replaced by discrete point charges, whose simple potential and field formulas 
 explain the  higher accuracy and speed of this method. We implemented the Gaussian
cubature method for the purpose of BEM computations both with CPU and GPU, and we compare its
performance with two different analytical integration methods.
The ten different Gaussian cubature formulas presented in our paper can be used for arbitrary high-precision 
and fast integrations over triangles and rectangles.
\end{abstract}

\medskip

\newpage

\tableofcontents

\section{Introduction}
 \label{SectionIntroduction}

Triangles and rectangles have many applications in mathematics, engineering and science.
The finite element method \cite{Huebner,Segerlind,Cook,Gupta} (FEM) and the boundary element method 
\cite{Gupta,KytheBEM,Gaul,Beer,Brebbia,BrebbiaDominguez,TJmaster} (BEM)
 rely especially heavily on triangles and rectangles
as basic elements for their discretization procedure. In order to obtain the system of equations with the nodal function values of
the elements in FEM and BEM, numerical or analytical integrations over the elements are needed.

The main subject of our paper is 3-dimensional electric potential and field calculation of 
charged triangular and rectangular BEM elements.
Electric field calculation is important in many areas of physics: electron and 
ion optics, charged particle beams, charged particle traps, electron microscopy, electron spectroscopy,
plasma and ion sources, electron guns, etc.
\cite{Szilagyi,HawkesKasper}.
 A special kind of electron and ion
energy spectroscopy is realized by the MAC-E filter spectrometers, where the integral
energy spectrum is measured by the combination of electrostatic retardation and magnetic
adiabatic collimation. Examples are the Mainz and Troitsk electron spectrometers
\cite{Mainz,Troitsk}, the aSPECT
proton spectrometer \cite{aSPECT}, and
the KATRIN pre- and main electron spectrometers  \cite{KATRIN,Prall,Arenz}.
High accuracy electric field and potential computations are indispensable for precise and reliable
charged particle tracking calculations for these experiments. For this purpose 
the open source C++ codes {\it KEMField} \cite{TJdiss} and {\it Kassiopeia} \cite{Furse,Groh,Grohdiss}
are used in the KATRIN and aSPECT experiments.

For the purpose of electric potential and field computation with BEM \cite{Szilagyi,HawkesKasper,ChariSalon,TJmaster},
the surface of the electrodes is discretized by many small boundary elements, and
a linear algebraic equation system is obtained for the unknown charge densities of the individual elements.
To solve these equations, either a direct or an iterative method is used.
When the charge densities are known, the potential and field at an arbitrary point (called field point) can be
computed by summing the potential and field contributions of all elements.
The electric potential and field of a single charged element at the field point can be expressed by analytical or 
numerical integration of the point-charge Coulomb formula 
times the charge density over the element surface.
In the simplest case of constant BEM element, the charge density 
is assumed to be constant over the element surface.

It is a general belief that analytical integration is more accurate than numerical integration. This is
in many cases true, but not always. A simple example with one-dimensional integral and 
small integration interval (Sec. \ref{SectionAnalytical}) shows that analytical integration can have a
large rounding error, due to the finite arithmetic precision of the computer \cite{Ueberhuber,Goldberg},
 while a simple numerical integration has in this case a much higher accuracy.
Analytically calculated electric potential and field formulas of charged triangles and rectangles
show a similar behavior for large distance ratios; the latter is defined as the distance of the element 
center and the field point divided by the average side length of the element: for field points far from the
element the distance ratio is large.
Sec. \ref{SectionAnalytical} presents a few plots for the relative error of the analytical potential and
field of triangles and rectangles as a function of the distance ratio. One can see that
for field points far from the element the
analytically computed potential and field values have significant rounding errors that are much 
higher than the machine epsilon of the corresponding arithmetic precision \cite{Ueberhuber,Goldberg} 
that is used for the computation. The analytical integration formulas of Refs.
\cite{TJdiss,Formaggio} and \cite{Hanninen,Hilkdiss} were used for these plots. 
There are many other published analytical
integration results for potential and field calculation of triangles and rectangles 
\cite{Rao,Okon,Davey,Tatematsu,Mukhopadhyay,Lopez,Carley,Durand,Birtles,Eupper}, and these suffer 
probably from the same rounding error that increases with the distance ratio.
The rounding errors are caused by the transcendental functions (e.g. log, atan2 etc.) in the analytical
formulas and by the well-known subtraction cancellation problem of finite-digit 
floating-point arithmetic computations.
Another disadvantage of the
analytical integrations is that the analytical potential and field formulas are rather complicated, 
and consequently the  calculations are rather time consuming.

According to the aforementioned simple example with one-dimensional integration, one  
can eliminate the rounding error problem of the analytical potential and field calculations by using
numerical integration. The two-dimensional numerical integration could be performed by two subsequent
one-dimensional integrations, using e.g. the efficient Gauss-Legendre quadrature method with 16 nodes
for each dimension \cite{Evans,Kythe}. If the field point is not too close to the triangle or rectangle,
 this bi-quadrature method results in very accurate integral values. In fact, we use this method as a reference
integration in order to define the errors of the other integration methods. Nevertheless, if the goal is also to
minimize the computation time, then it is more expedient to use the Gaussian cubature method for
two-dimensional numerical integration, because it has fewer function evaluations for a targeted accuracy.
In Sec. \ref{SectionCubature} we present a short overview about the numerical integration of an arbitrary
function over a triangle or a rectangle with Gaussian cubature. 
The integral is approximated by a weighted sum of the function values at a given number of Gaussian points.
The accuracy of this approximation is defined by the degree of the cubature formula, which usually
increases with the number $N$ of Gaussian points.
Appendices A and B contain ten tables of the 
Gaussian points and weights of five different Gaussian cubature formulas for triangles and rectangles. These
formulas have various number of Gaussian points (from $N=4$ to $N=33$) and degrees of accuracy (from 3 to 13).

In Sec. \ref{SectionPotentialField} we apply  the general Gaussian cubature formulas for electric potential
and field calculation of triangles and rectangles with constant charge density.
The mathematical formalism of the Gaussian cubature integration is illustrated by a nice physical picture:
the triangles and rectangles with continuous charge distribution are replaced by discrete point charges,
and the complicated two-dimensional integration of the potential-field calculation is substantially simplified
by using the potential and field formulas of point charges. The figures in Sec. \ref{SectionPotentialField} 
reveal that for field points far from the elements the Gaussian cubature method has high accuracy and is
exempt from the rounding error problem of the analytical integrations. The relative error of the potential
and field calculation with a given Gaussian cubature formula is about  $10^{-15}$ above some distance ratio
limit (with double precision computer arithmetics),
 and increases with decreasing distance ratio below this limit, which decreases with increasing
number of Gaussian points (i.e. for more accurate cubature formulas). For small distance ratios, e.g. 
${\rm DR}<3$, the Gaussian cubature formulas presented in our paper
 are not accurate enough, and analytical integration should be
used in this region.

Sec. \ref{SectionManyElements} contains results for accuracy comparisons of potential and field simulations
with two complex electrode geometries containing 1.5 million triangles and 1.5 million rectangles, respectively.
The table in this section shows that the relative errors of the Gaussian cubature calculations are much
smaller than the errors of the analytical calculations. In fact, the  Gaussian cubature relative error values are close
to $10^{-15}$, i.e. the best accuracy that is possible to obtain with the double precision arithmetic that is used
in these computations. Sec. \ref{SectionCompTime} presents an additional advantage of the Gaussian cubature method: the potential
and field calculations with low-$N$ cubature formulas are significantly faster than the corresponding simulations
with analytical integration. E.g. the 7-point cubature method in CPU computations is about five times faster than the analytical
integration method of Ref. \cite{Hanninen,Hilkdiss} and more than ten times faster than the analytical
integration method of Ref. \cite{TJdiss}. The larger speed of the Gaussian cubature method is due to its
simplicity: it needs the evaluation of only one square root and one division operation for each Gaussian point, in addition to
multiplications and additions, while the analytical integration methods need many time consuming
transcendental function evaluations (like log, atan2 etc.). 
Of course, for the elements with smaller distance ratios the cubature formulas with larger number of points
(e.g. 12, 19 or 33) have to be used (to get an acceptable accuracy level), and they are slower
(although still faster than the analytical methods).
Nevertheless, the distance ratio distribution plots in Sec. \ref{SectionManyElements} show that  
for most of the elements of a typical electrode geometry 
the fast 7-point cubature method can be used.

\section{Analytical integration}
 \label{SectionAnalytical}

Let us first consider the simple one-dimensional integral of the function $\exp(x)$ from 1 to $1+\delta$.
The analytical result from the Newton-Leibniz formula is $\exp(1+\delta) - \exp(1)$, and 
this is for small $\delta$ a typical example for loss of accuracy (digits) when subtracting
two almost equal numbers. The problem is that the computer stores in a
floating-point arithmetic number always a finite number of digits \cite{Ueberhuber,Goldberg}, 
and a few of them can disappear at subtraction.
Therefore, the relative error of the above integral for small $\delta$ is much larger than the 
machine epsilon of the floating-point number system that is used for the computation.

If a high accuracy is required for the integral value with small $\delta$, it is better to use numerical integration,
e.g. Gaussian quadrature. With double precision arithmetic  and with 
Gauss-Legendre quadrature \cite{Evans,Kythe}  using 16 x 16 integration nodes, 
the relative error of the integral is smaller than $10^{-15}$, i.e. the numerical integration has
the maximal precision that is possible to achieve with the corresponding floating-point number system.
One can see this by the independence of the integral value on the number of integration nodes, or by comparing with
a higher precision (e.g. long double) computation. Due to this fact, we can get the accuracy of the analytical integral values by comparing
them with the numerical quadrature values: ${\rm err}(I_{an})=|(I_{an}-I_{num})/I_{num}|$.

\begin{figure}[!htb]
    \centering
    \includegraphics[width=0.65\textwidth]{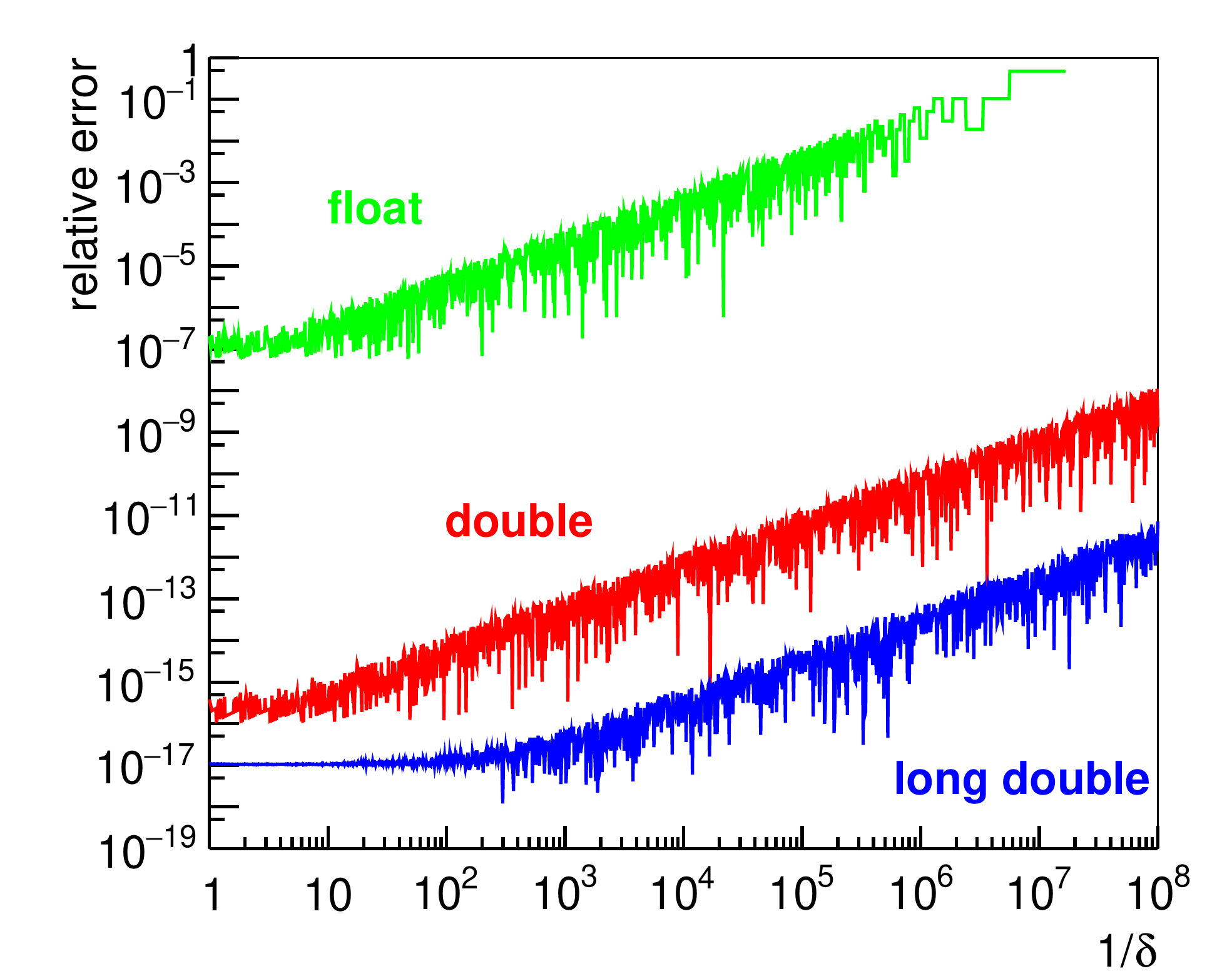}
    \caption{Relative error of the one-dimensional analytical integral $\exp(1+\delta) - \exp(1)$ as function of $1/\delta$, for  float, double and long double C++ arithmetic types.}
    \label{FigExpTest}
\end{figure}

Fig. \ref{FigExpTest} presents the relative error  ${\rm err}(I_{an})$
of the analytical integral above as a function of $1/\delta$, 
for three different C++ floating-point arithmetic types: float, double and long double.
The relative error increases with $1/\delta$ and decreases with increasing precision of the floating-point
arithmetic. It is then obvious that the analytical integration has a rounding error which can be for small
 integration interval size $\delta$ much larger than the precision of the floating-point arithmetic type that is used
for the calculation.
The numerical integration (like Gaussian quadrature) is, however, devoid of this precision loss
problem.

Let us continue now with electric potential and field computation of triangles and rectangles with constant
charge density $\sigma$. The potential $\Phi$ and field ${\bf E}$ at field point ${\bf P}$
can be generally written (in SI units) as
surface integrals over the element:

\begin{equation}
    \label{Eqpotfield}
    \Phi({\bf P})=\frac{\sigma}{4\pi\varepsilon_0} \int\limits_{\rm element}d^2 {\bf Q} \cdot 
    \frac{1}{|{\bf P}-{\bf Q}|},   \quad \quad
    {\bf E}({\bf P})=\frac{\sigma}{4\pi\varepsilon_0} \int\limits_{\rm element}d^2 {\bf Q} \cdot 
    \frac{{\bf P}-{\bf Q}}{|{\bf P}-{\bf Q}|^3},
\end{equation}
where ${\bf Q}$ denotes the integration point on the element surface.
We introduce  the distance $D$ between the
field point ${\bf P}$ and the center point (centroid) ${\bf Q_{\rm cen}}$ of the triangle or rectangle:
$D=|{\bf P}-{\bf Q_{\rm cen}}|$, and the average side length $L$ of the triangle or rectangle
(e.g. $L=(a+b+c)/3$, with triangle side lengths $a$, $b$ and $c$). Then, the distance ratio of the
element and field point combination is defined
as ${\rm DR}=D/L$; this corresponds to the $1/\delta$ parameter of the one-dimensional integral described above.

\begin{figure}[htbp]
    \centering
    \subfigure[TrianglePotential]{\includegraphics[width=0.48\textwidth]{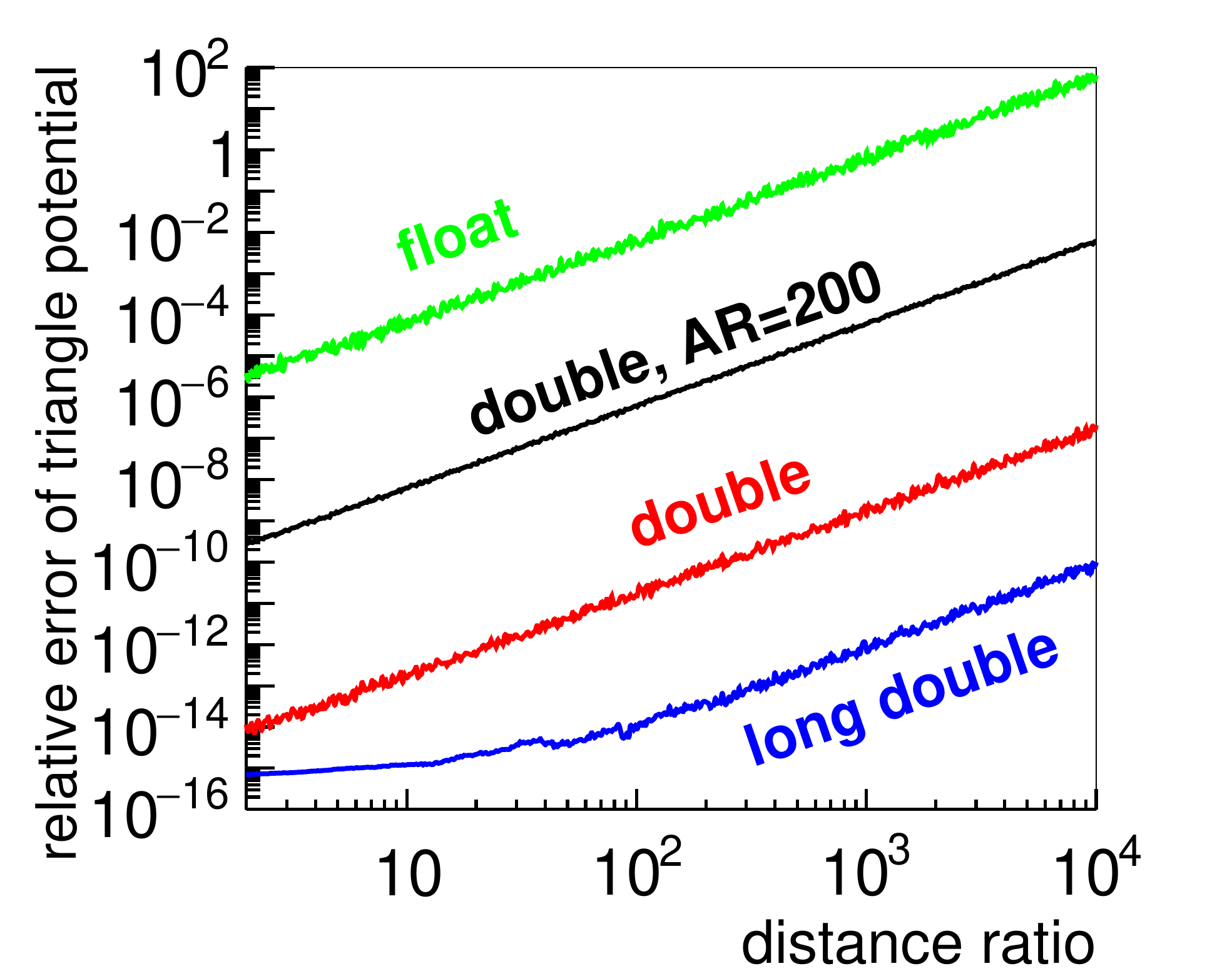}\label{FigTrianglePotTJ}}\quad
    \subfigure[TriangleField]{\includegraphics[width=0.48\textwidth]{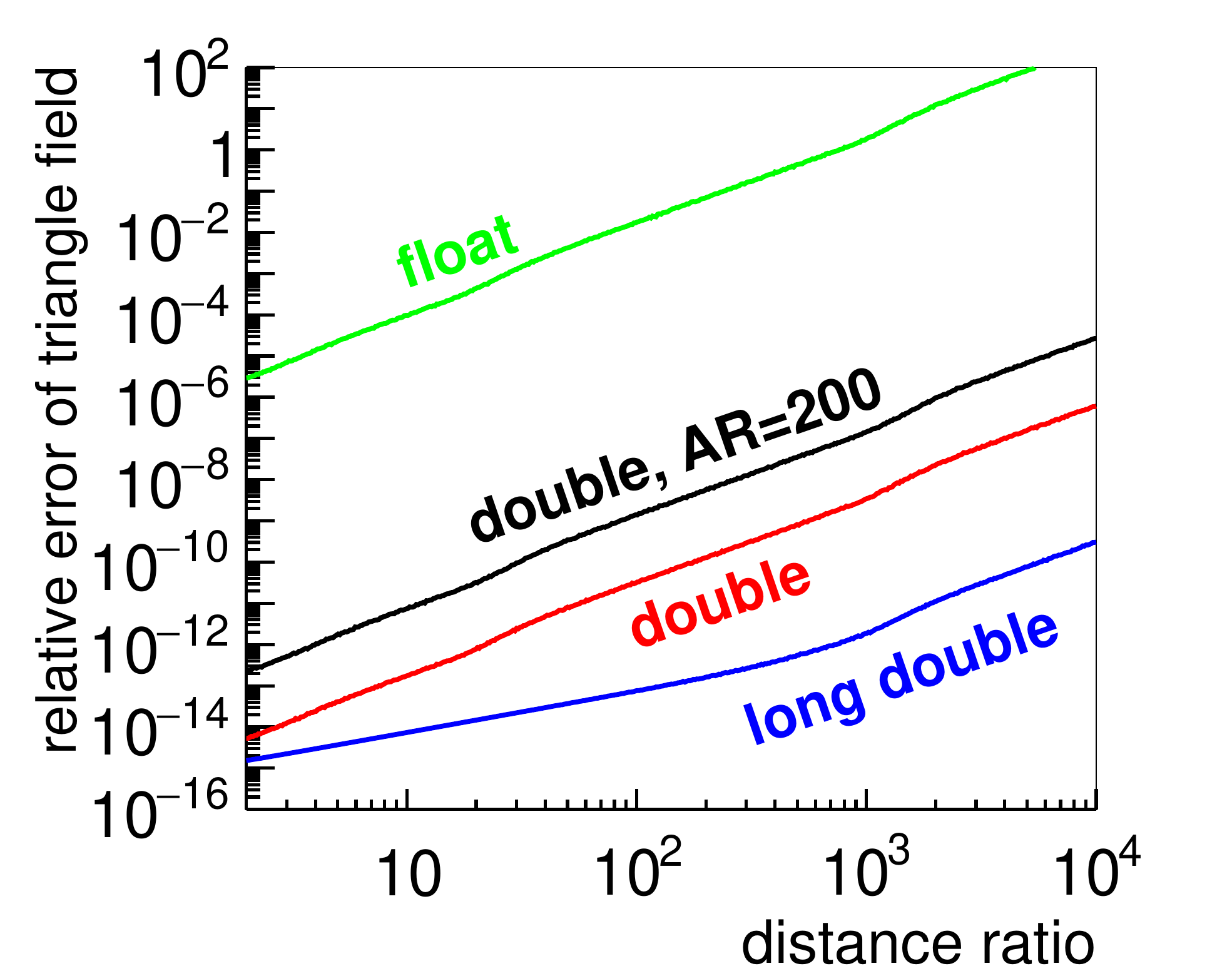}\label{FigTriangleFieldRWG}}         
    \caption{Averaged relative error of the analytically computed  triangle potential
    (Refs. \cite{TJdiss,Formaggio}, left) and triangle field (Refs. \cite{Hanninen,Hilkdiss}, right),
    as a function of the distance ratio, for  float (green), double (red) and long double (blue) C++ arithmetic types
    (with low aspect ratio triangles),
    and with double precision and high aspect ratio (AR=200) triangles (black).}
    \label{FigTriangleAnal}
\end{figure}

\begin{figure}[htbp]
    \centering
    \subfigure[RectanglePotential]{\includegraphics[width=0.48\textwidth]{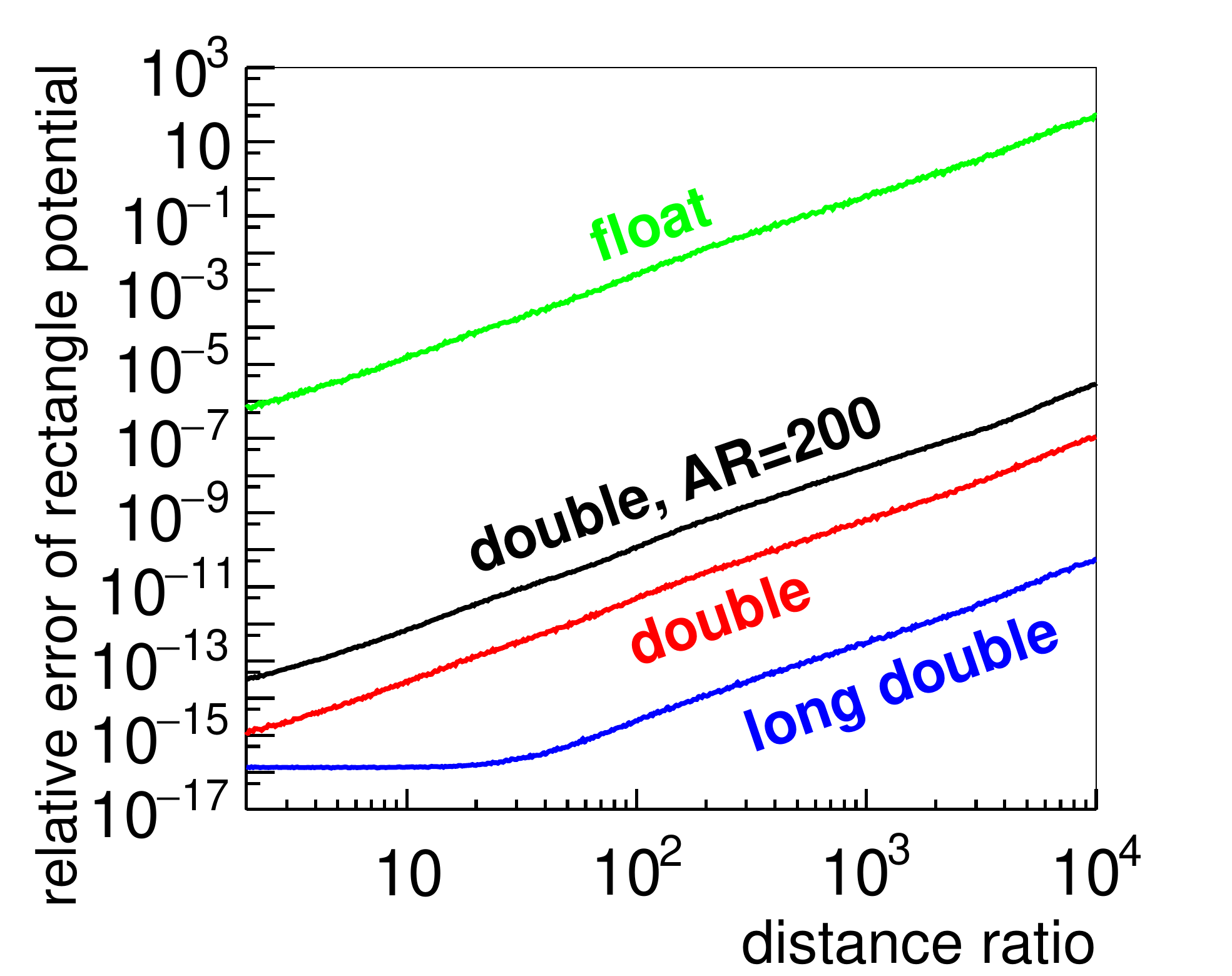}\label{FigRectanglePotRWGH}}\quad
    \subfigure[RectangleField]{\includegraphics[width=0.48\textwidth]{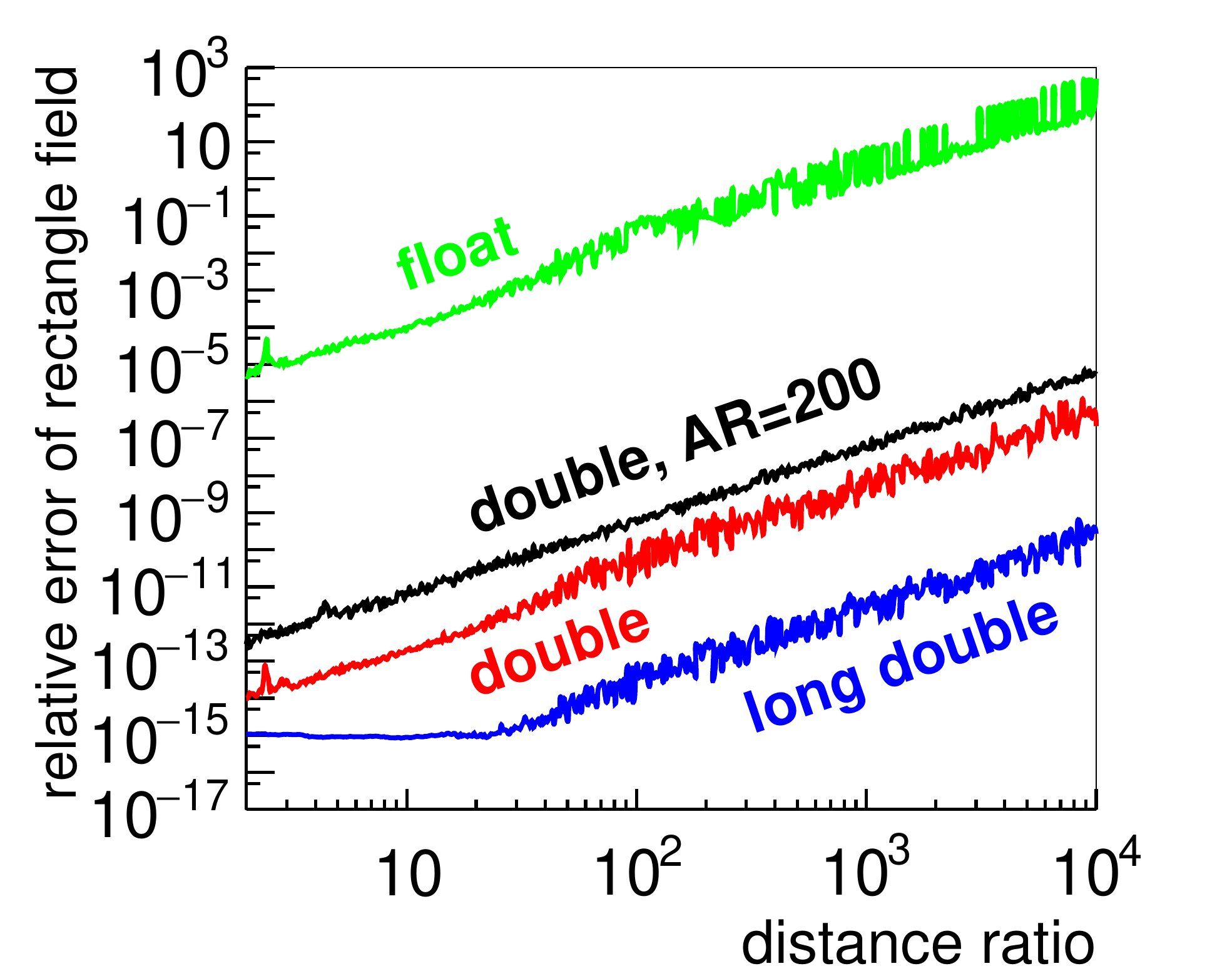}\label{FigRectangleFieldTJH}}         
    \caption{Averaged relative error of the analytically computed  rectangle potential
    (Refs. \cite{Hanninen,Hilkdiss}, left) and rectangle field (Ref. \cite{TJdiss}, right),
    as a function of the distance ratio, for  float (green), double (red) and long double (blue) C++ arithmetic types
    (with low aspect ratio rectangles),
    and with double precision and high aspect ratio (AR=200) rectangles (black).}
    \label{FigRectangleAnal}
\end{figure}

In order to investigate the potential and field calculation of triangles and rectangles, we carried out the 
following procedure. First, we generated  the corner points of triangles or rectangles randomly inside a cube
with unit lengths and the direction unit vector of the field point also randomly relative to the triangle or rectangle
centroid. Then, for a fixed distance ratio both the element (triangle or rectangle) and the field point are
defined. We computed the potential and field of the element with unit charge density at the field point by two
different analytical integration methods: first, 
with Refs. \cite{TJdiss} (App. A and B) and \cite{Formaggio} (App. A), and second, with Refs. \cite{Hanninen}
(Eqs. 63 and 74) and \cite{Hilkdiss}.
In order to obtain the error of the analytical integrals, we also calculated the potential and field by numerical
integration, using two successive
one-dimensional Gauss-Legendre quadratures (GL2) \cite{Evans,Kythe}
with $n_{GL2}=16$ integration nodes for both integrations. If the field
point is located not too close to the element (i.e. for distance ratio above two), the latter method yields a relative accuracy
which is close to the precision of the applied floating-point arithmetic type (e.g. order of $10^{-15}$ for
double precision in C++). One can check that the GL2 integral values do not change if the
discretization number $n_{GL2}$ is changed, and possible rounding errors can be tested by
comparing double and long double calculations.
The relative error of the analytically computed potential is then:
${\rm err}(\Phi_{an})=|(\Phi_{an}-\Phi_{GL2})/\Phi_{GL2}|$. For the field ${\bf E}$ we define the relative error
in the following way: ${\rm err}({\bf E}_{an})=\sum_{j=x,y,z} |{\bf E}_{j,an}-{\bf E}_{j,GL2}|/|{\bf  E}|_{GL2}$, where
the sum goes over the  components $x$, $y$ and $z$. 
We generate 1000 elements and field point direction vectors for each distance ratio,
 and we take the average
of the above defined relative error values. In this case the triangles and rectangles have small (mainly below ten)
aspect ratios (AR). The triangle aspect ratio is defined as the longest side length divided by the corresponding 
height. Similarly, the rectangle aspect ratio is the longer side  divided by the shorter side.

\begin{figure}[htbp]
    \centering
    \subfigure[Triangles]{\includegraphics[width=0.48\textwidth]{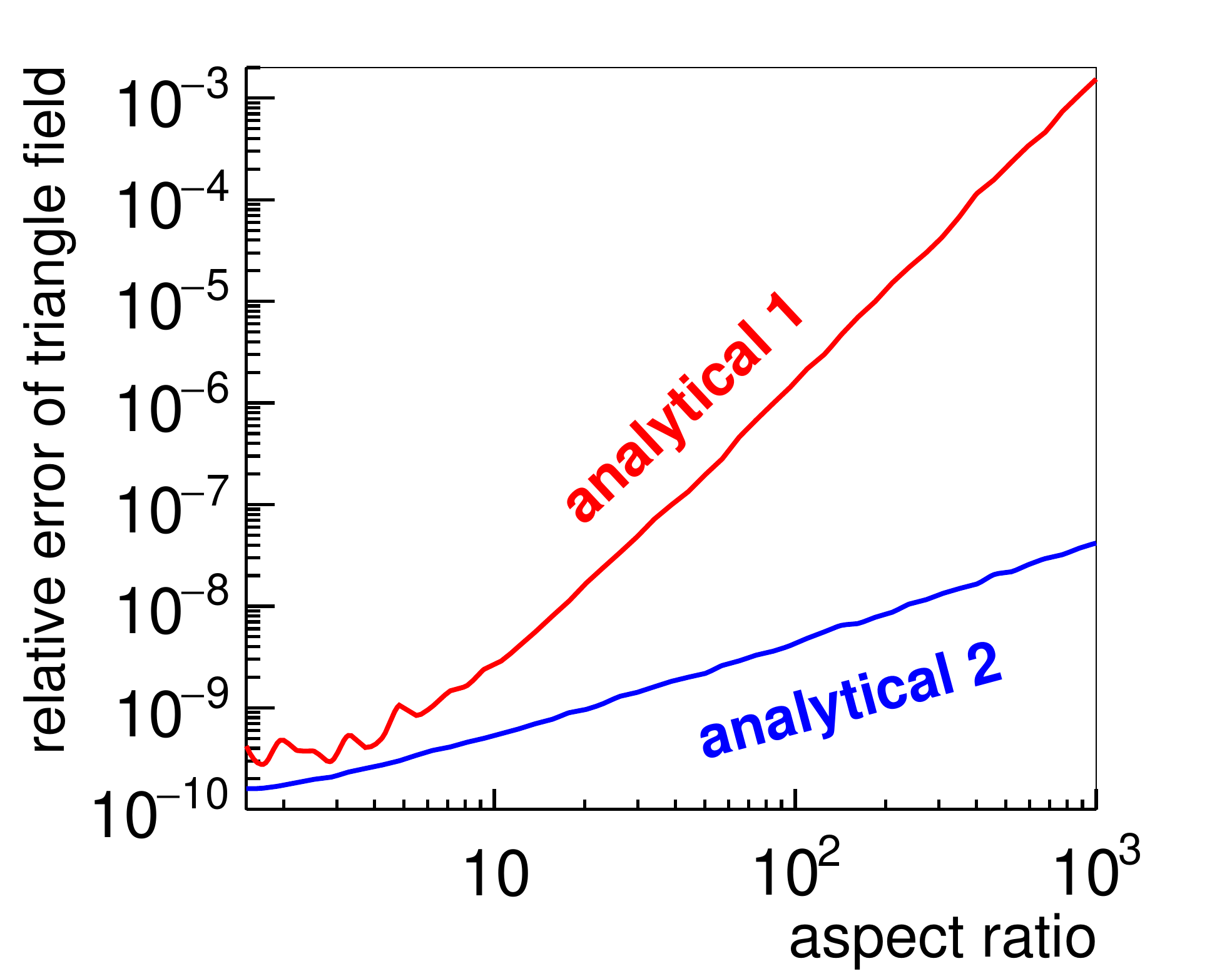}\label{FigARTriangle}}\quad
    \subfigure[Rectangles]{\includegraphics[width=0.48\textwidth]{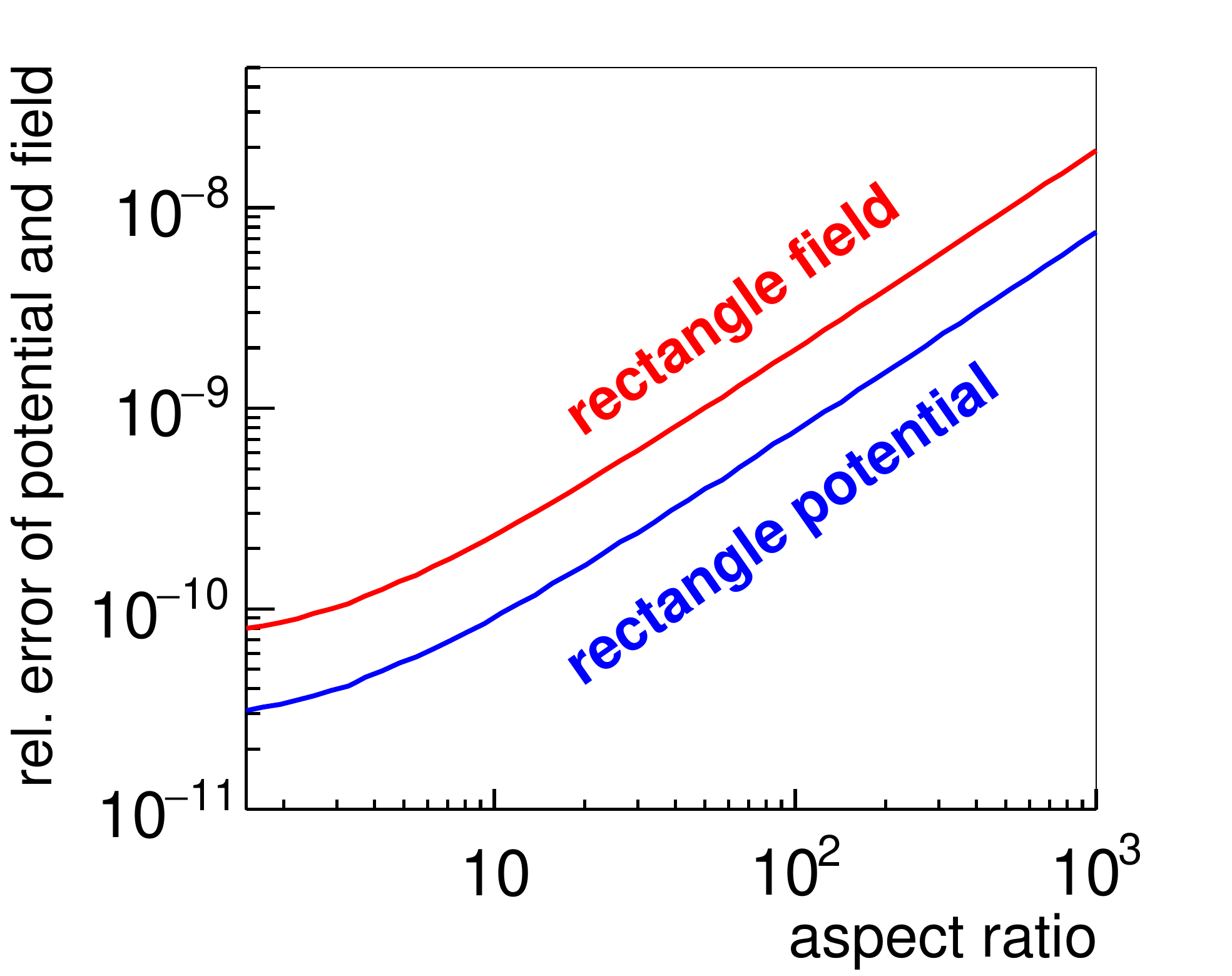}\label{FigARRectangle}}         
    \caption{Averaged relative error of the analytically computed  triangle field
    (left, 1 (red): Ref.  \cite{TJdiss}, 2 (blue): Ref. \cite{Hanninen}), and rectangle potential and field 
    (right, Ref. \cite{Hanninen}; Ref.  \cite{TJdiss} is similar),
    as a function of the aspect ratio, with fixed distance ratio DR=300.}
    \label{FigARAnalytical}
\end{figure}

Figures \ref{FigTriangleAnal} and \ref{FigRectangleAnal} present the above defined averaged relative errors
for triangle and rectangle potential and field, computed by the analytical integration formulas of Refs.
\cite{TJdiss,Formaggio,Hanninen,Hilkdiss}, as a function of the distance ratio, for three different floating-point 
arithmetic types (float, double and long double of the C++ language) with low aspect ratio elements,
 and also for larger (AR=200) aspect ratio elements. 
For small distance ratio (below five), the relative error values are close to the corresponding floating-point
arithmetic precision. Note that our Gauss-Legendre quadrature implementation has double precision accuracy,
therefore the relative error in case of long double precision is not smaller than $10^{-15}$ or so.
Furthermore, the plots show that the relative error of the analytical integrals increases with the distance ratio 
and also with the aspect ratio, while it
decreases with increasing precision of the floating-point arithmetic type.

It is obvious that the analytical integrals have significant rounding errors for large distance ratio.
These potential and field errors are even much larger for triangles and 
rectangles with large aspect ratio, as one can see in Fig. \ref{FigARAnalytical}.
We conjecture that all other analytically integrated potential and field formulas in the literature (Refs. 
\cite{Rao,Okon,Davey,Tatematsu,Mukhopadhyay,Lopez,Carley,Durand,Birtles,Eupper})
 have similarly large rounding errors for field points located far from the elements.

\section{Numerical integration with Gaussian cubature}
 \label{SectionCubature}

The rounding error problem of the analytical integration can be solved by using numerical integration
for field points far away from the element. Gaussian quadrature and cubature are efficient numerical integration
techniques \cite{Evans,Kythe,Stroud,Engels,EngelnMullges}
which can yield high accuracies with a minimal number of nodes (function evaluation points).

The integral of an arbitrary function $f$ over a surface element can be generally approximated by Gaussian cubature as

\begin{equation}  \label{Eqcubatureformula}
    \int\limits_{\rm element}d^2{\bf Q}  \cdot f({\bf Q}) = 
    {\cal A} \cdot \sum_{i=1}^N w_i \cdot f({\bf Q}_i) +{\cal R}, \quad \quad
    \sum_{i=1}^N w_i = 1,
\end{equation}
where ${\bf Q}_i$ and $w_i$ are the Gaussian points (nodes, knots) and weights, respectively,
 and ${\cal A}$ denotes the
area of the element. 
The remainder ${\cal R}$ is the absolute error of the Gaussian cubature integral formula.

\begin{figure}[htbp]
    \centering
    \subfigure[Rectangle]{\includegraphics[width=0.47\textwidth]{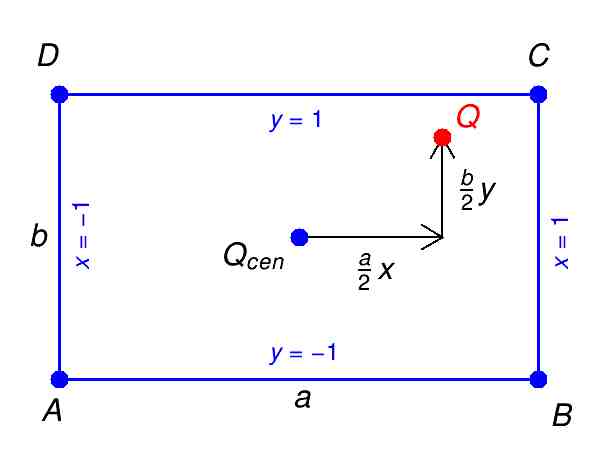}\label{FigRectangleQ}}\quad
    \subfigure[Triangle]{\includegraphics[width=0.47\textwidth]{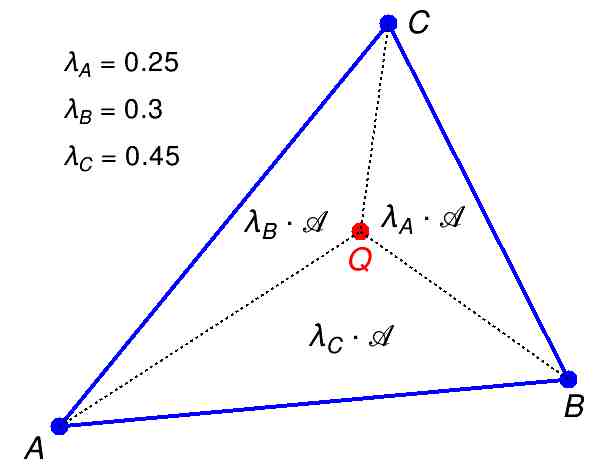}\label{FigTriangleQ}}         
    \caption{Natural local coordinates on a rectangle (left) and barycentric (area) coordinates on a triangle
    (right).}
    \label{FigRectangleTriangleQ}
\end{figure}

To parametrize the Gaussian points ${\bf Q}_i$, it is expedient to use local coordinates:
they rely on the element geometry for their definition and
 are generally called natural coordinates \cite{Huebner,Cook,Segerlind}.
In the case of rectangles, it is advantageous to use a local coordinate system whose axes are
parallel with the side unit vectors ${\bf u}_x$ and ${\bf u}_y$ of the rectangle. 
An arbitrary point ${\bf Q}$ on the plane of the rectangle can be parametrized by the local natural 
coordinates $x$ and $y$:

\begin{equation}  \label{EqrectangleQ}
    {\bf Q}= {\bf Q}_{\rm cen} +\frac{a}{2}\, x\, {\bf u}_x + \frac{b}{2}\, y\, {\bf u}_y, 
\end{equation}
where ${\bf Q}_{\rm cen}$ denotes the rectangle center, and $a$ and $b$ are the two side lengths;
see Fig. \ref{FigRectangleQ}. For $|x|\le 1$ and $|y|\le 1$ the point ${\bf Q}$ is inside the rectangle, 
otherwise it is outside.

For the parametrization of points in a triangle it is advantageous to use the so-called
barycentric or area coordinates. 
An arbitrary point ${\bf Q}$ on the plane of the triangle can be written as a linear combination 
of the triangle vertex vectors ${\bf A}$, ${\bf B}$ and ${\bf C}$:

\begin{equation}  \label{Eqbarycentric}
    {\bf Q}=\lambda_A {\bf A} + \lambda_B {\bf B} +  \lambda_C {\bf C}, \quad 
    {\rm with} \quad  \lambda_A + \lambda_B + \lambda_C =1,
\end{equation}
where  $\lambda_A $, $\lambda_B$ and $ \lambda_C$ are the barycentric coordinates. They are all less than 1 if the
point ${\bf Q}$ is inside the triangle. For $\lambda_A=1, \; \lambda_B=\lambda_C=0$:  ${\bf Q}={\bf A}$, 
and for $\lambda_A=0$  the point 
${\bf Q}$ is on the line ${\bf BC}$. $\lambda_A=\lambda_B=\lambda_C=1/3$ corresponds to the centroid of the triangle.
The coordinate $\lambda_A$ is equal to the ratio of the triangle areas
${\bf QBC}$ and  ${\bf ABC}$ (and similarly for $\lambda_B$ and $\lambda_C$), as one can see
in Fig. \ref{FigTriangleQ}. Due to this property,  the barycentric coordinates are also
called area coordinates (see Refs. \cite{Huebner,Segerlind,Cook}).

The computation time of a Gaussian cubature formula is proportional to the number of 
Gaussian points (nodes) $N$.
A good  formula has small error (remainder ${\cal R}$ in Eq. \ref{Eqcubatureformula})
with small $N$. Usually, a  two-dimensional Gaussian cubature formula is constructed so that
it is exact (with ${\cal R}=0$) for all possible monomials $f(x,y)=x^n y^m$ with $0\le n+m\le d$,
but for $n+m>d$ the remainder ${\cal R}$ is not zero. The integer $d$ is called the degree
of the cubature formula. A large degree $d$ corresponds to high accuracy, but the number of nodes $N$, 
and so the computation time, also increases with $d$.

The definition of the degree above makes it plausible how to determine the nodes and weights for 
two-dimensional Gaussian cubature formulas: first, one calculates the integral
 (analytically or numerically) on the left-hand side of Eq.
\ref{Eqcubatureformula} for several different monomial functions
$f(x,y)=x^n y^m$ (with $n+m \le d$). Then, each integral value is set equal to the cubature sum
formula on the right-hand side of Eq. \ref{Eqcubatureformula} (with ${\cal R}=0$).
One obtains then the nodes and weights by solving this nonlinear equation system,
which is obviously a difficult task, especially for large $N$ and $d$.
It is expedient to have all weights positive (to reduce rounding errors) and all nodes inside the element.

Gaussian point coordinates  and weights  for triangles and rectangles with various
$N$ and $d$ values can be found in several books  \cite{Stroud,Engels,EngelnMullges}
and in many publications 
\cite{Radon,HammerStroud1956,HammerStroud1958,HammerMarloweStroud,
   Gatermann,LynessJespersen,Papanicolopulos,AlbrechtCollatz,Tyler,Moller,CoolsHaegemans,
 Rabinowitz,Omelyan,Dunavant,HaegemansPiessens1976,HaegemansPiessens1977,LaursenGellert,
 Wandzura,Zhang}. Examples for very high degree cubature formulas are: $N=175,\, d=30$ for triangles
in \cite{Wandzura}, and $N=100,\, d=23$ for rectangles in \cite{Omelyan}.
There are  several review papers about the subject in the literature 
\cite{CoolsRabinowitz,LynessCools,Cools1999,Cools2002,Cools2003,Coolswebpage}.

In App. \ref{AppendixTriangles} we present barycentric coordinates and weights of five different 
Gaussian cubature formulas 
for triangles: $N=4,\, d=3$ (Table \ref{TableTriangle4}), $N=7,\, d=5$ (Table \ref{TableTriangle7}),
$N=12,\, d=7$ (Table \ref{TableTriangle12}), $N=19,\, d=9$ (Table \ref{TableTriangle19}) and
$N=33,\, d=12$ (Table \ref{TableTriangle33}).
App. \ref{AppendixRectangles}  contains cartesian natural coordinates and weights 
of five different Gaussian cubature formulas 
for rectangles: $N=4,\, d=3$ (Table \ref{TableRectangle4}), $N=7,\, d=5$ (Table \ref{TableRectangle7}),
$N=12,\, d=7$ (Table \ref{TableRectangle12}), $N=17,\, d=9$ (Table \ref{TableRectangle17}) and
$N=33,\, d=13$ (Table \ref{TableRectangle33}). In most cases, one row in a table corresponds to
several nodes with equal weights: the coordinates of the other nodes can be obtained 
by various permutations or sign changes of the given numbers (see the table captions for detailed
explanations).  

\begin{figure}[htbp]
    \centering
    \subfigure[Triangle7]{\includegraphics[width=0.47\textwidth]{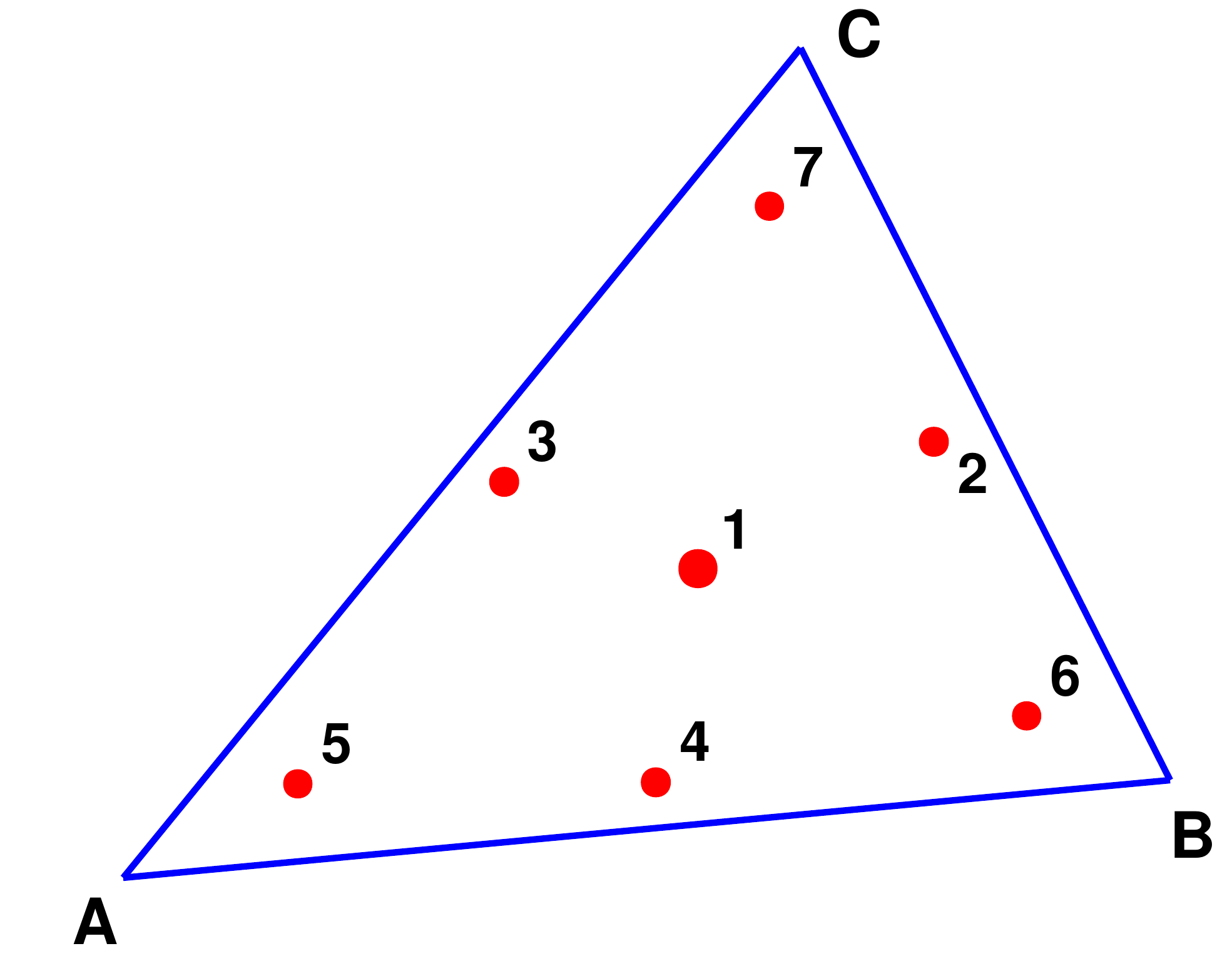}\label{FigTriangle7}}\quad
    \subfigure[Triangle12]{\includegraphics[width=0.47\textwidth]{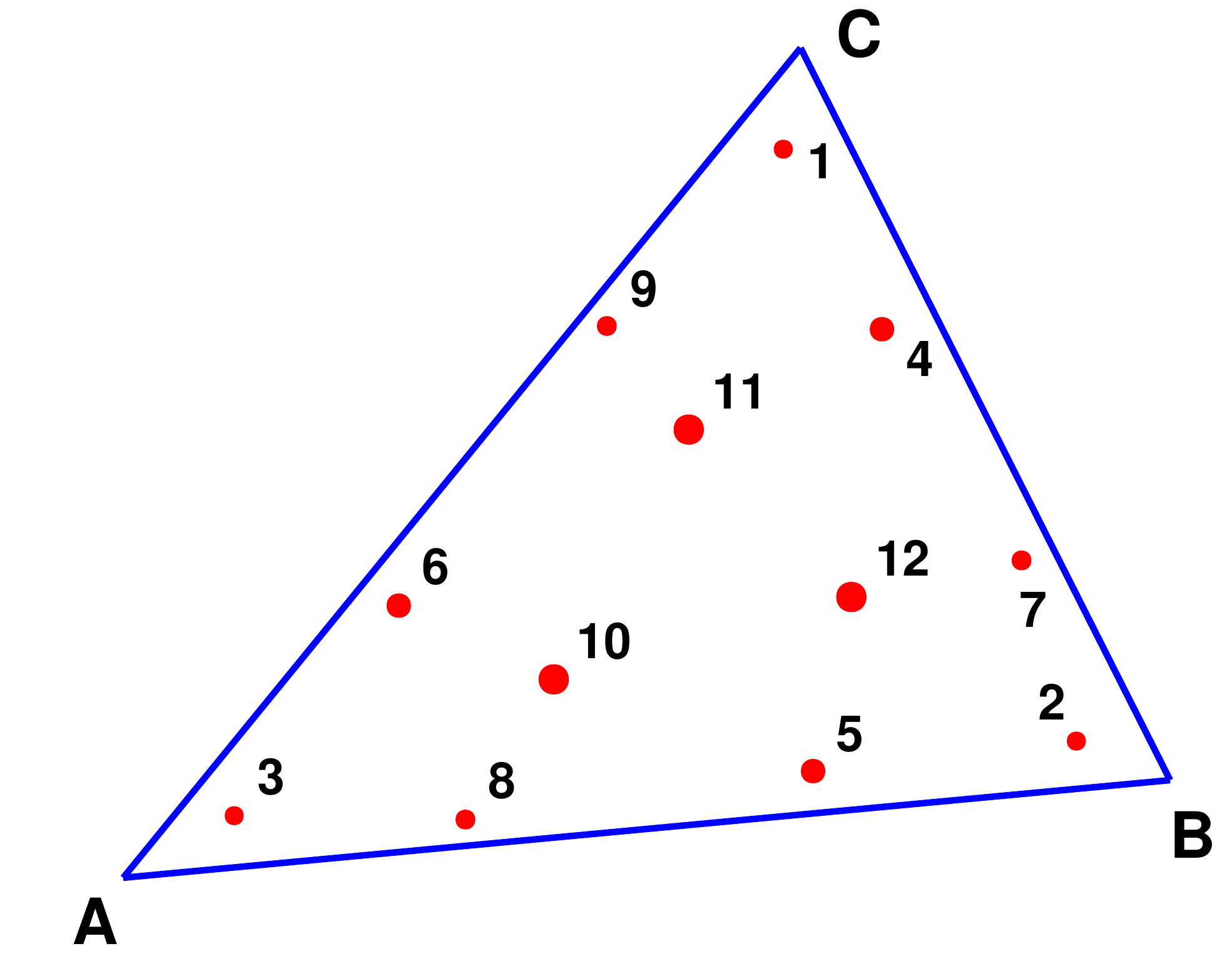}\label{FigTriangle12}}         
    \caption{Gaussian points of the 7-point (left) and the 12-point (right) cubature formula for triangle.
    The numbers are the indices of the Gaussian points. The surfaces of the red circles are proportional to the
    corresponding weights.}
    \label{FigTriangle}
\end{figure}

\begin{figure}[htbp]
    \centering
    \subfigure[Rectangle7]{\includegraphics[width=0.47\textwidth]{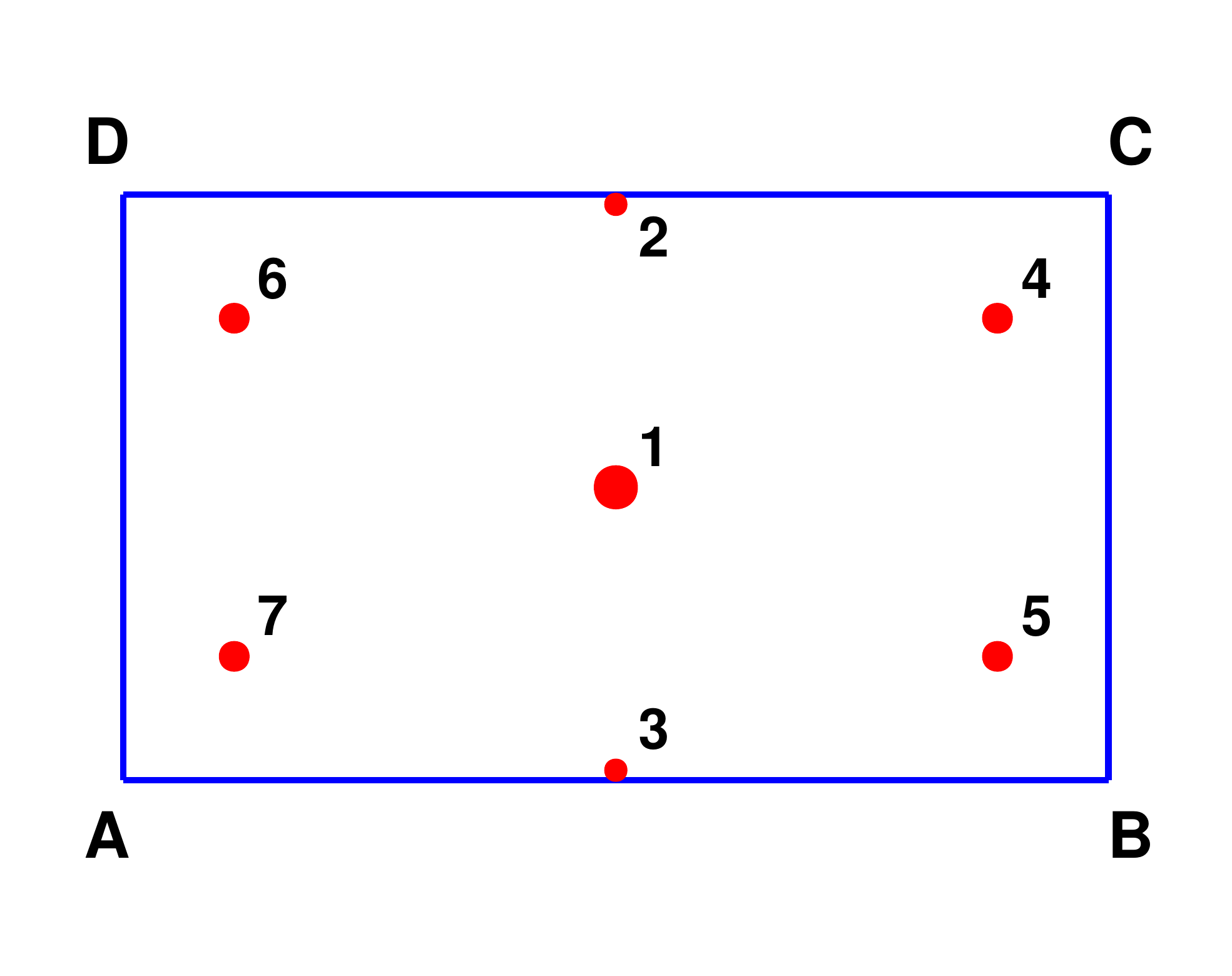}\label{FigRectangle7}}\quad
    \subfigure[Rectangle17]{\includegraphics[width=0.47\textwidth]{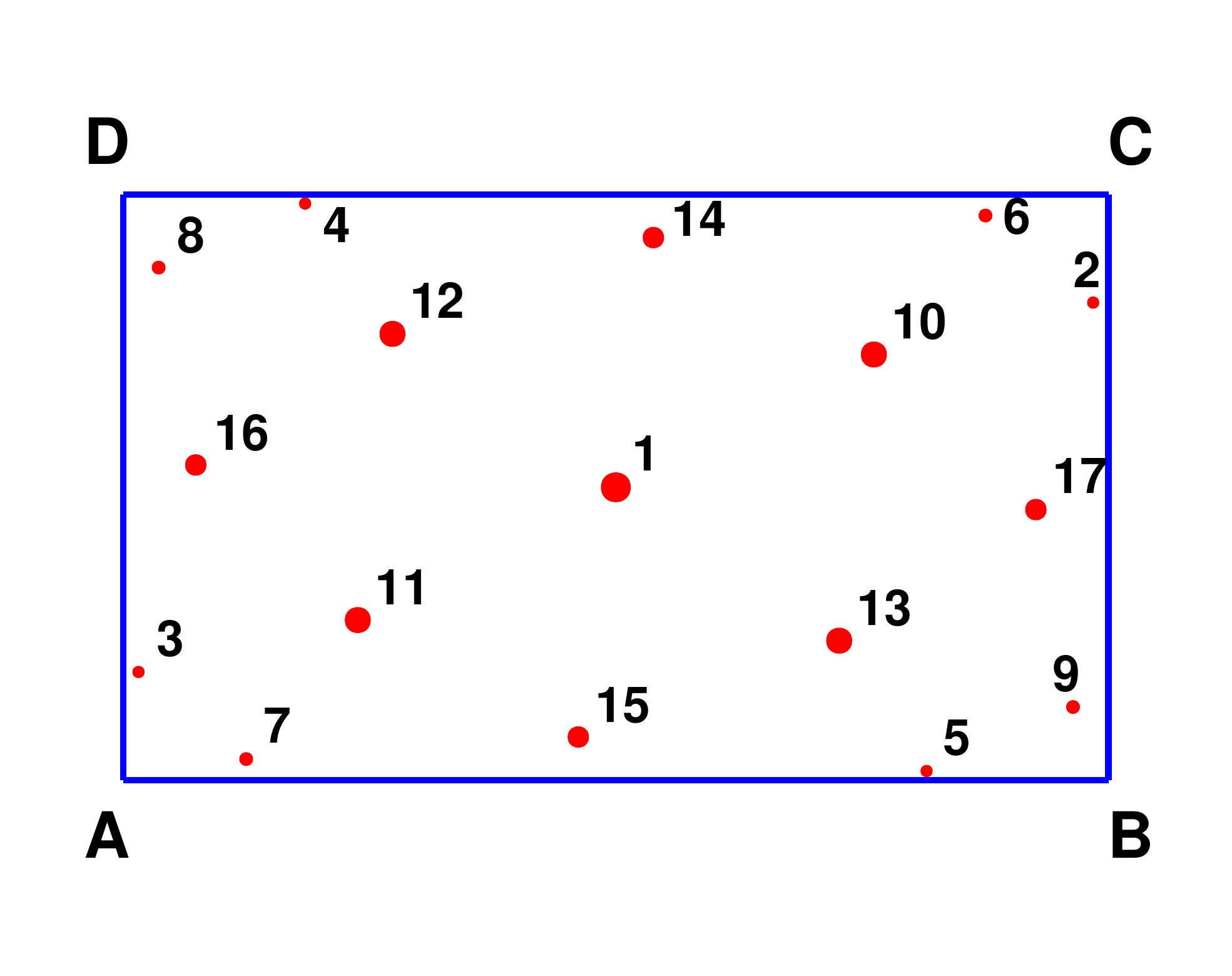}\label{FigRectangle17}}         
    \caption{Gaussian points of the 7-point (left) and the 17-point (right) cubature formula for rectangle.
    The numbers are the indices of the Gaussian points. The surfaces of the red circles are proportional to the
    corresponding weights.}
    \label{FigRectangle}
\end{figure}

Figures \ref{FigTriangle} and \ref{FigRectangle} show a few examples for the Gaussian points of a
triangle and a rectangle. In Fig. \ref{FigTriangle7}, point 1 corresponds to the first row in Table
\ref{TableTriangle7} (this is the centroid of the triangle). Points 2 to 4 correspond to the second row:
for point 2 $\to \lambda_A=t+2ts$, for point 3 $\to \lambda_B=t+2ts$, and for point 4 $\to \lambda_C=t+2ts$
(the other barycentric coordinates are $t-ts$). Similarly, points 5 to 7 correspond to the third row in that table.
These figures can be useful to understand the multiplicity structure of the tables in App. 
\ref{AppendixTriangles} and \ref{AppendixRectangles}.

\section{Potential and field calculation by Gaussian cubature}
 \label{SectionPotentialField}

The electric potential and field of an arbitrary constant BEM element (with constant charge density $\sigma$)
at a field point ${\bf P}$ can be
approximated by Gaussian cubature as:
\begin{equation}  \label{Eqpotfieldcubature}
    \Phi({\bf P}) \approx \frac{1}{4\pi\varepsilon_0} 
    \sum_{i=1}^N \frac{q_i}{|{\bf P}-{\bf Q}_i|},
    \quad \quad
    {\bf E}({\bf P}) \approx  \frac{1}{4\pi\varepsilon_0} 
            \sum_{i=1}^N  q_i   \frac{{\bf P}-{\bf Q}_i}{|{\bf P}-{\bf Q}_i|^3},
\end{equation}
where $q_i=w_i q$ denotes the charge of point $i$, 
and $q=\sigma {\cal A}$ is the total charge of the element. 
We have then a nice and intuitive physical picture of the Gaussian cubature
formalism: the BEM element with continuous charge density is replaced by discrete point charges, whereas
the charge $q_i$ at the Gaussian point ${\bf Q}_i$ is proportional to the Gaussian weight $w_i$,
and the sum of the individual charges $q_i$ is equal to the total charge of the element
(see Eq. \ref{Eqcubatureformula}).
Obviously, it is much more easier to compute the potential and field produced by point charges instead of
continuous charge distributions. As we will see below, the point charge method is not only easier 
but also more precise,
at least for field points which are not too close to the element. In addition, the point charge calculation
is typically faster than the analytical integration with continuous charge distribution
(see Sec. \ref{SectionCompTime}).

\begin{figure}[htbp]
    \centering
    \subfigure[TrianglePotential]{\includegraphics[width=0.48\textwidth]
   {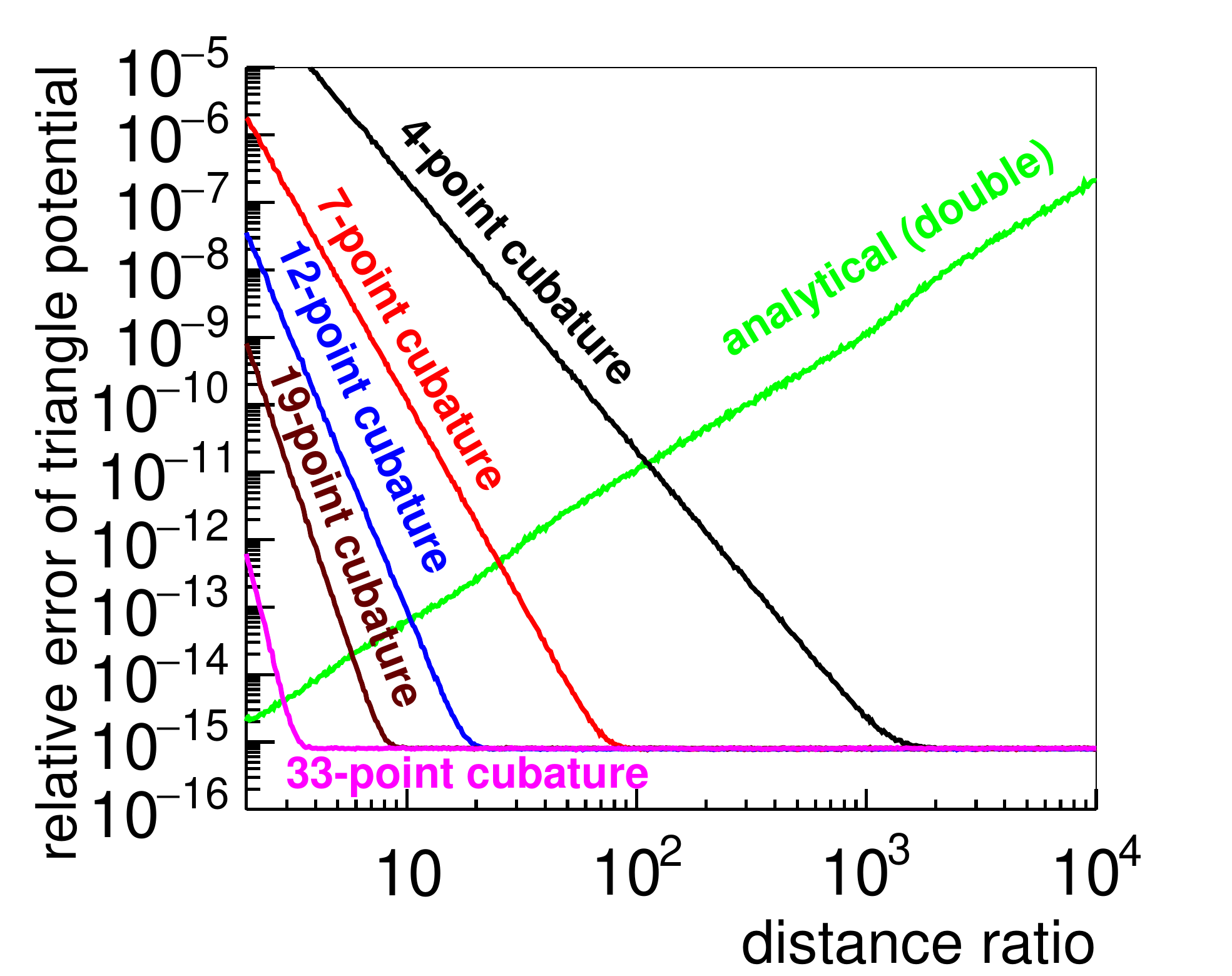}\label{FigTrianglePotRWGCubature}}\quad
    \subfigure[TriangleField]{\includegraphics[width=0.48\textwidth]
   {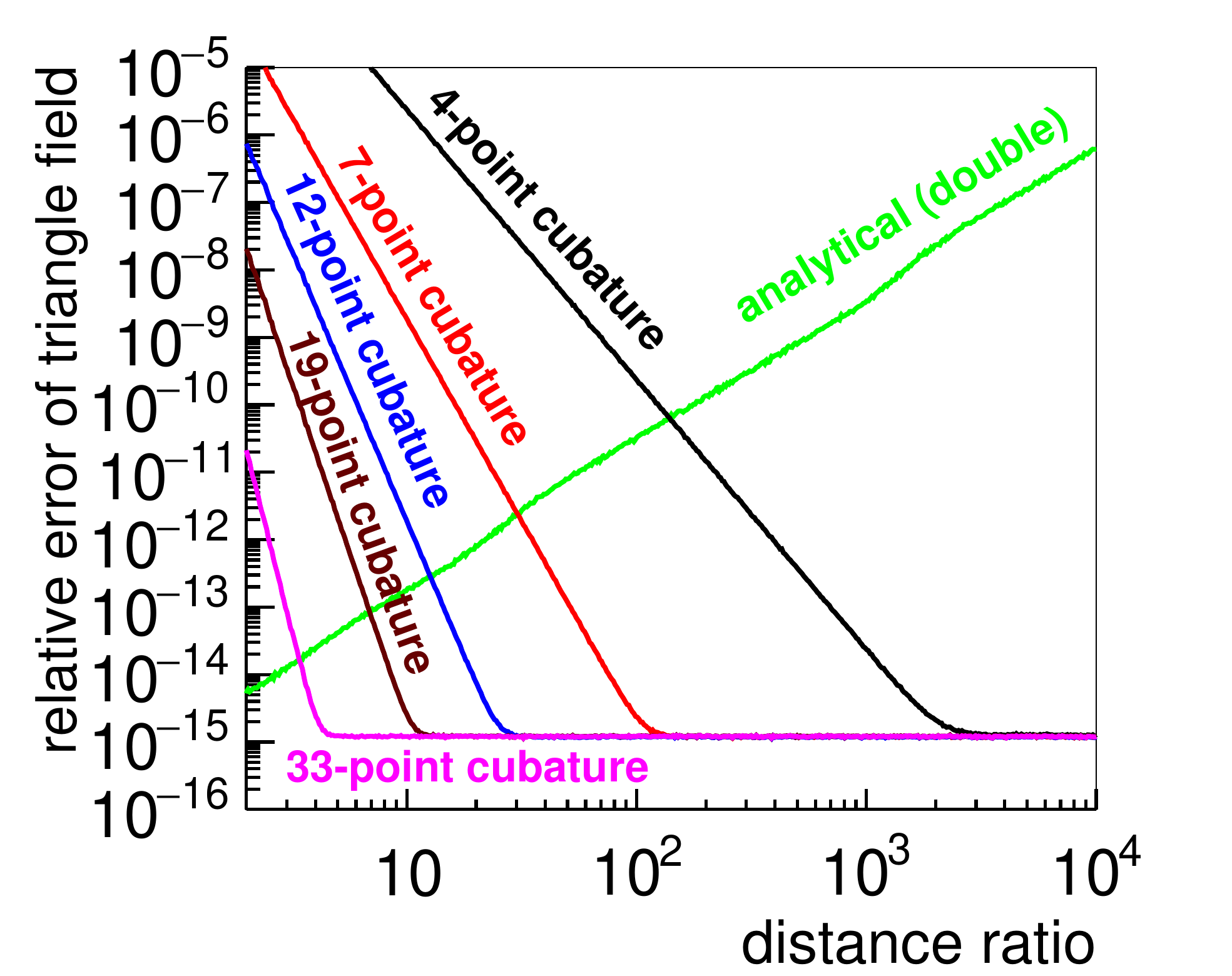}\label{FigTriangleFieldRWGCubature}}         
    \caption{Averaged relative error of the triangle potential (left) and field (right) for the 5  Gaussian cubature
    approximations of App. \ref{AppendixTriangles}, as a function of the distance ratio.
    For comparison, the averaged relative error of the triangle potential and field computed by analytical integration 
    (Refs. \cite{Hanninen,Hilkdiss}) is also shown.}
    \label{FigTriangleRWGCubature}
\end{figure}

\begin{figure}[htbp]
    \centering
    \subfigure[RectanglePotential]{\includegraphics[width=0.48\textwidth]
   {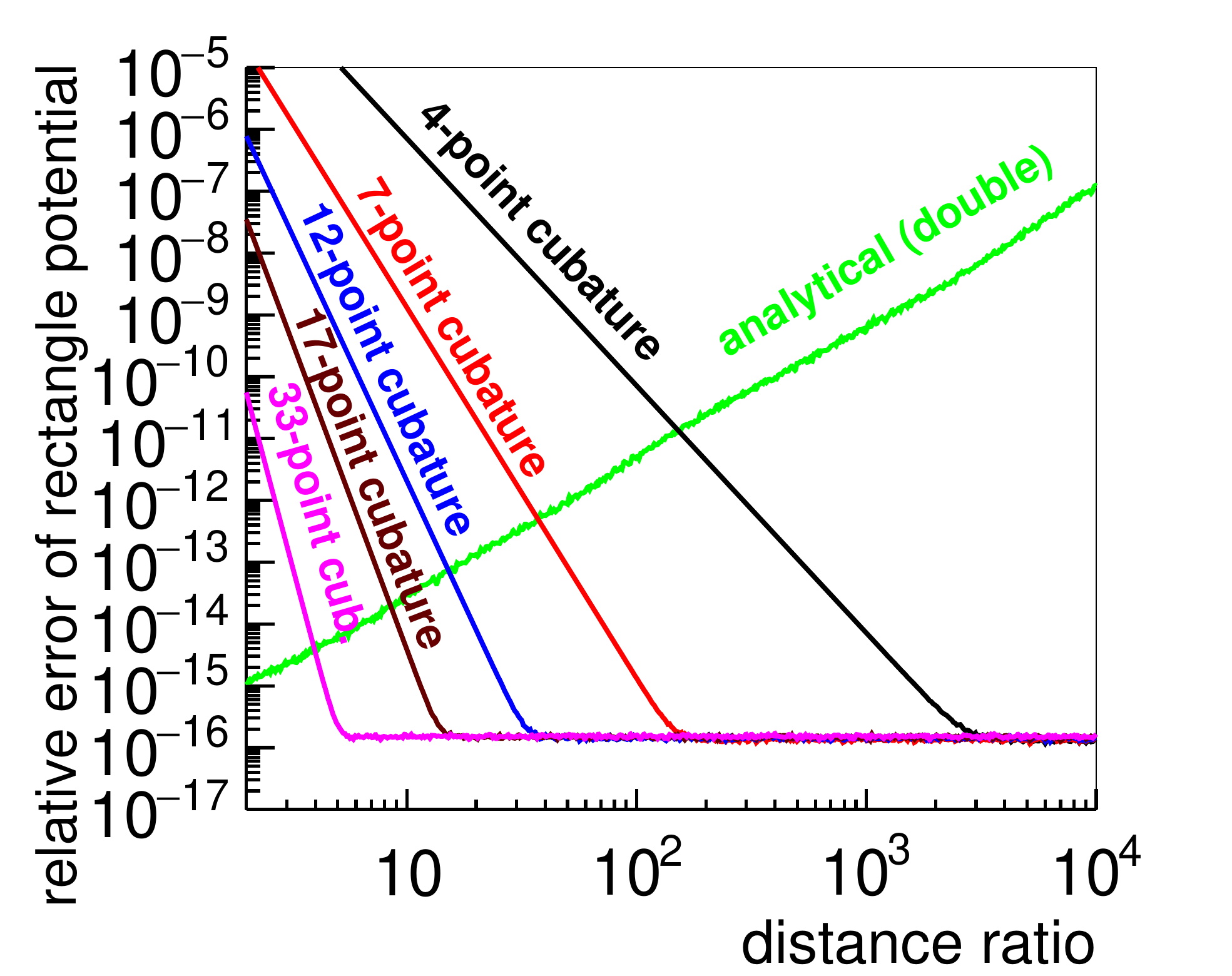}\label{FigRectanglePotRWGCubature}}\quad
    \subfigure[RectangleField]{\includegraphics[width=0.48\textwidth]
   {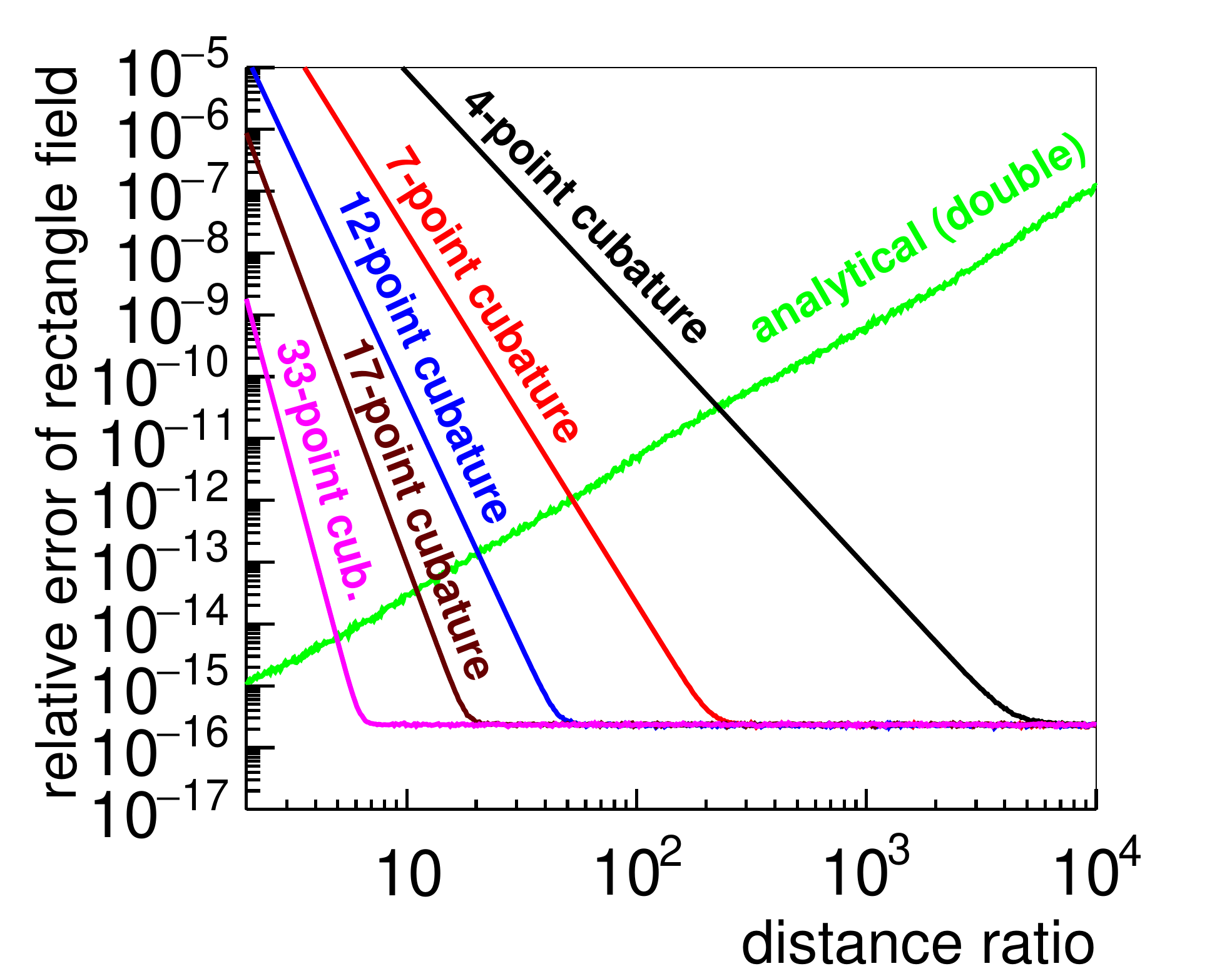}\label{FigRectangleFieldRWGCubature}}         
    \caption{Averaged relative error of the rectangle potential (left) and field (right) for the 5  Gaussian cubature
    approximations of App. \ref{AppendixRectangles}, as a function of the distance ratio.
    For comparison, the averaged relative error of the rectangle potential and field computed by analytical integration 
    (Refs. \cite{Hanninen,Hilkdiss}) is also shown.}
    \label{FigRectangleRWGCubature}
\end{figure}

In order to investigate the performance of the Gaussian cubature or point charge approximation for
potential and field calculation of triangles and rectangles, we used the same procedure that is described
in Sec. \ref{SectionAnalytical}. First, for a fixed element and field point, the relative error of the
potential computed by Gaussian cubature is defined as:
${\rm err}(\Phi_{cub})=|(\Phi_{cub}-\Phi_{GL2})/\Phi_{GL2}|$, where $\Phi_{GL2}$ is computed by
two-dimensional  Gauss-Legendre integration. A similar formula holds for the field error
(see in Sec. \ref{SectionAnalytical}). Then, 1000 elements and field point
directions are randomly generated for a fixed distance ratio, and the averages of the relative error values
are calculated for 500 different distance ratio values from 2 to 10000.

Figures \ref{FigTriangleRWGCubature} and \ref{FigRectangleRWGCubature}
present the averaged relative error
of the potential and field for triangles and rectangles as a function of the distance ratio, for the ten Gaussian cubature
approximations described in App. \ref{AppendixTriangles} and \ref{AppendixRectangles}, together with the
relative error of the analytical integration described in Refs. \cite{Hanninen,Hilkdiss} 
(with  double precision arithmetic type). One can see that, while the relative error of the analytical integration is
small for field points close to the element (small distance ratio), and it increases with the distance ratio, the
behavior of the Gaussian cubature error is just the opposite: it is large for field points near the element, and
it decreases with the distance ratio.
In fact, the accuracy of the Gaussian cubature  at high distance ratio is limited only by the 
finite-digit computer arithmetic precision.
Therefore, it seems that the analytical and numerical integration methods  complement each other:
to get high accuracy everywhere, one should use analytical integration for field points close to the
element and numerical integration farther away.

It is also conspicuous from the figures that the cubature
formulas with more Gaussian points $N$ have higher accuracy and can be used also for field points closer to the
elements to obtain a given accuracy level (e.g. $10^{-15}$). The computation time for the potential or field
simulation by Gaussian cubature is approximately
proportional to $N$, therefore, in order to minimize the computation
time, it is expedient to use several different cubature formulas: for large distance ratio 
DR one can use a cubature formula with smaller $N$, and for smaller DR one should use a formula
with larger $N$. E.g. to obtain $2\cdot10^{-15}$ relative accuracy level for the triangle potential, one should use
the following cubature formulas in the various distance ratio intervals:
$N=4$ for ${\rm DR}>1500$, $N=7$ for $80<{\rm DR}<1500$, $N=12$ for $20<{\rm DR}<80$,
$N=19$ for $8<{\rm DR}<20$, $N=33$ for $3<{\rm DR}<8$, and analytical integration for ${\rm DR}<3$.
In the case of triangle field these limits are slightly higher. 
If both the potential and the field has to be computed for a field point in one computation step, then one should use the limits
defined by the field; namely, in this case the same Gaussian points and weights can always be used
for both calculations.

\begin{figure}[!htbp]
    \centering
    \includegraphics[width=0.65\textwidth]{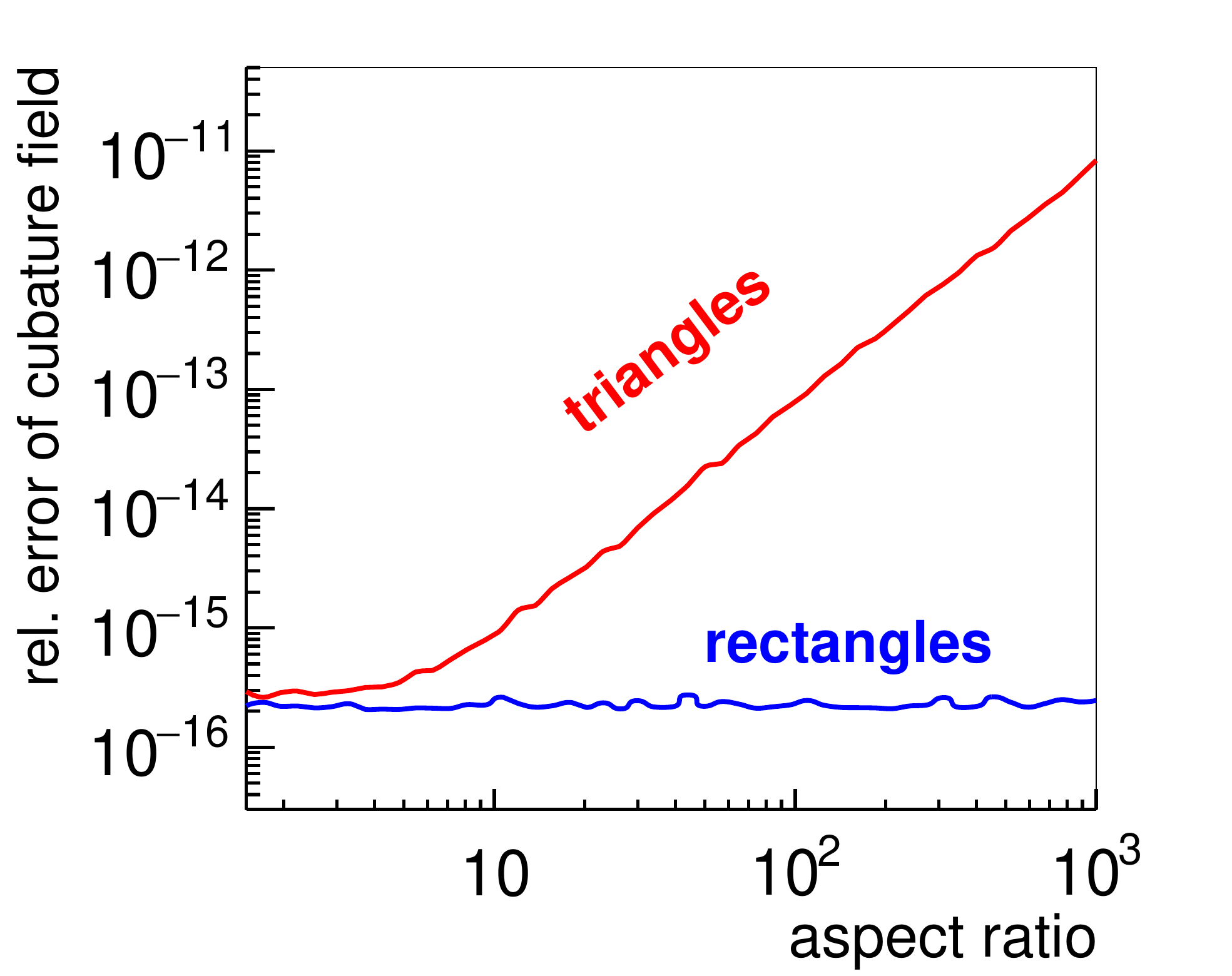}
    \caption{Averaged relative error of the 7-point  Gaussian cubature field for triangles and rectangles
    as a function of the aspect ratio, for a fixed distance ratio DR=300.}
    \label{FigARCubature}
\end{figure}

We showed in Sec. \ref{SectionAnalytical} that the relative error of the analytically computed potential
and field increases with the triangle and rectangle aspect ratio. We investigated the aspect 
ratio dependence also for Gaussian cubature. Fig. \ref{FigARCubature} presents the error of the 7-point
Gaussian cubature field as a function of the aspect ratio, for DR=300 distance ratio; the potential error
and higher-order Gaussian cubatures have a similar behavior. One can see that
the triangle potential and field error of the
Gaussian cubature increases with the aspect ratio. Therefore, a large number of triangles 
with  high aspect ratios should be
avoided in BEM calculations, if high accuracy computations are required. On the other hand, the
Gaussian cubature potential and field calculations of rectangles seem not to be sensitive to the 
rectangle aspect ratio.

Finally, we mention that a triangle can be mapped into a rectangle by the Duffy transformation
\cite{Duffy,LynessCools}, therefore the numerical integration over a triangle by Gaussian cubature
can also be done by using the Duffy transformation in conjunction with the Gaussian cubature formulas for rectangles
of App. \ref{AppendixRectangles}, instead of the triangle formulas of App. \ref{AppendixTriangles}.
In this case, however, we get much larger errors for the potential and field of triangles than by using
the Gaussian cubature formulas for triangles. E.g. the relative error of the triangle field at DR=100
with the Duffy transformation and the 7-point rectangle cubature formula is about $10^{-8}$,
in contrast with the few times $10^{-15}$ error of the triangle 7-point Gaussian cubature formula 
(see Fig. \ref{FigTriangleFieldRWGCubature}).

\section{Accuracy comparisons with complex geometries}
 \label{SectionManyElements}

\begin{figure}[!htbp]
    \centering
    \includegraphics[width=0.75\textwidth]{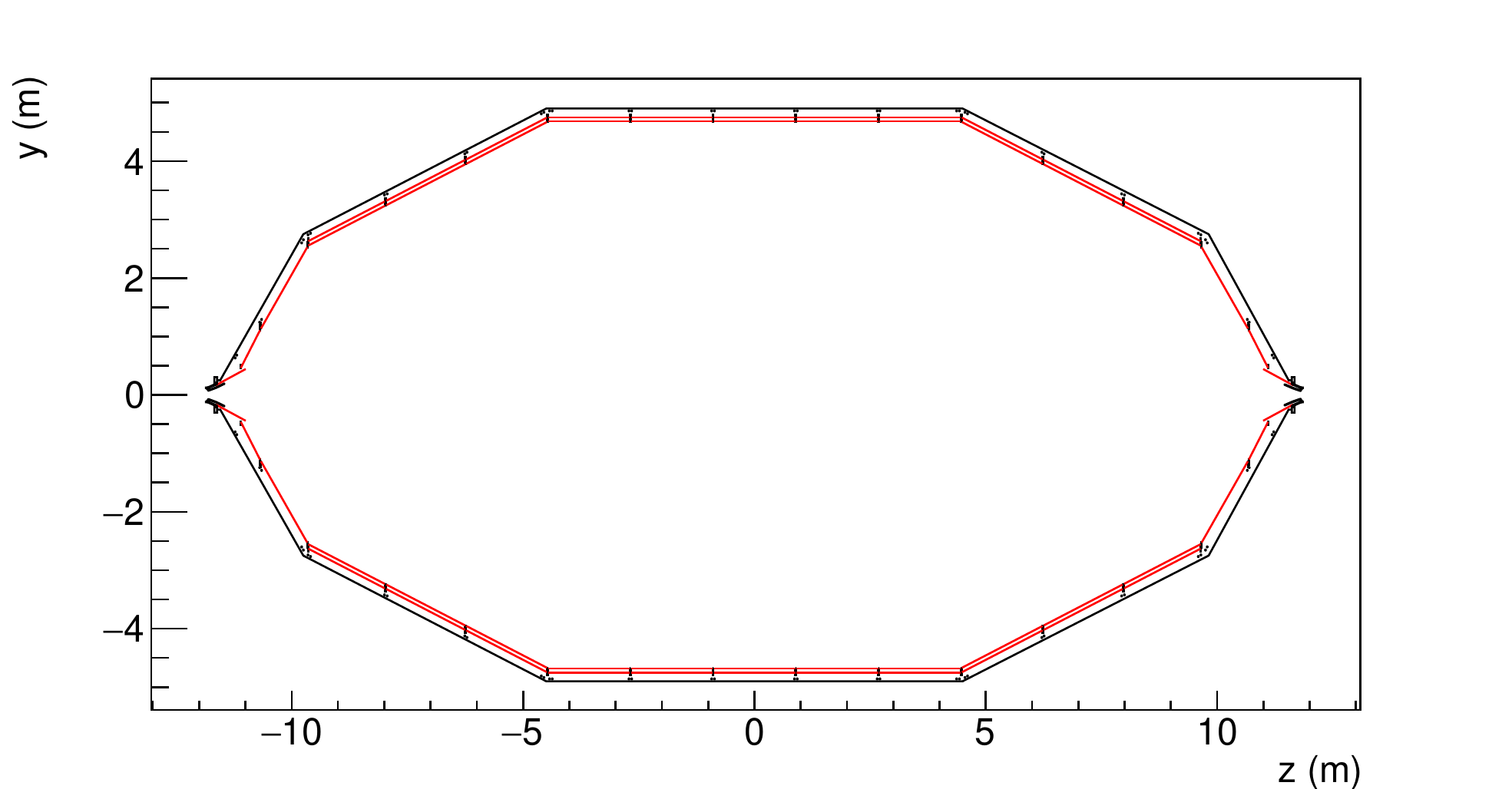}
    \caption{Schematic cross-sectional view of the KATRIN main spectrometer vessel 
(black, outside) and inner electrode system (red, inside).}
    \label{FigMainSpec}
\end{figure}

The main goal of potential and field calculation of charged triangles and rectangles is to apply these elements
for electric potential and field computations of complex electrode systems with BEM. Therefore, it is important
to compare the accuracy of the analytical and numerical integration methods not only for individual elements,
but also  for electrode systems with many elements. In this section, we present results for two 
electric field simulations: one of them contains only triangles as BEM elements, the other one only rectangles.
For this purpose, we discretized the main spectrometer vessel and inner electrodes of the KATRIN experiment
first by 1.5 million triangles and second by 1.5 million rectangles, using a dipole electrode potential configuration.
The KATRIN main spectrometer inner electrode system has a complicated wire electrode system
\cite{Valerius}
but in our models the wire electrodes are replaced by full electrode surfaces.
Fig. \ref{FigMainSpec} shows a two-dimensional plot (z-y plane) of the KATRIN main spectrometer vessel
and inner electrode system. For further details about this spectrometer we refer to Ref. \cite{Arenz}.
The dipole electrode model discretizations are explained in detail in Ref. \cite{Stern}.

\begin{figure}[htbp]
    \centering
    \subfigure[Triangles]{\includegraphics[width=0.48\textwidth]{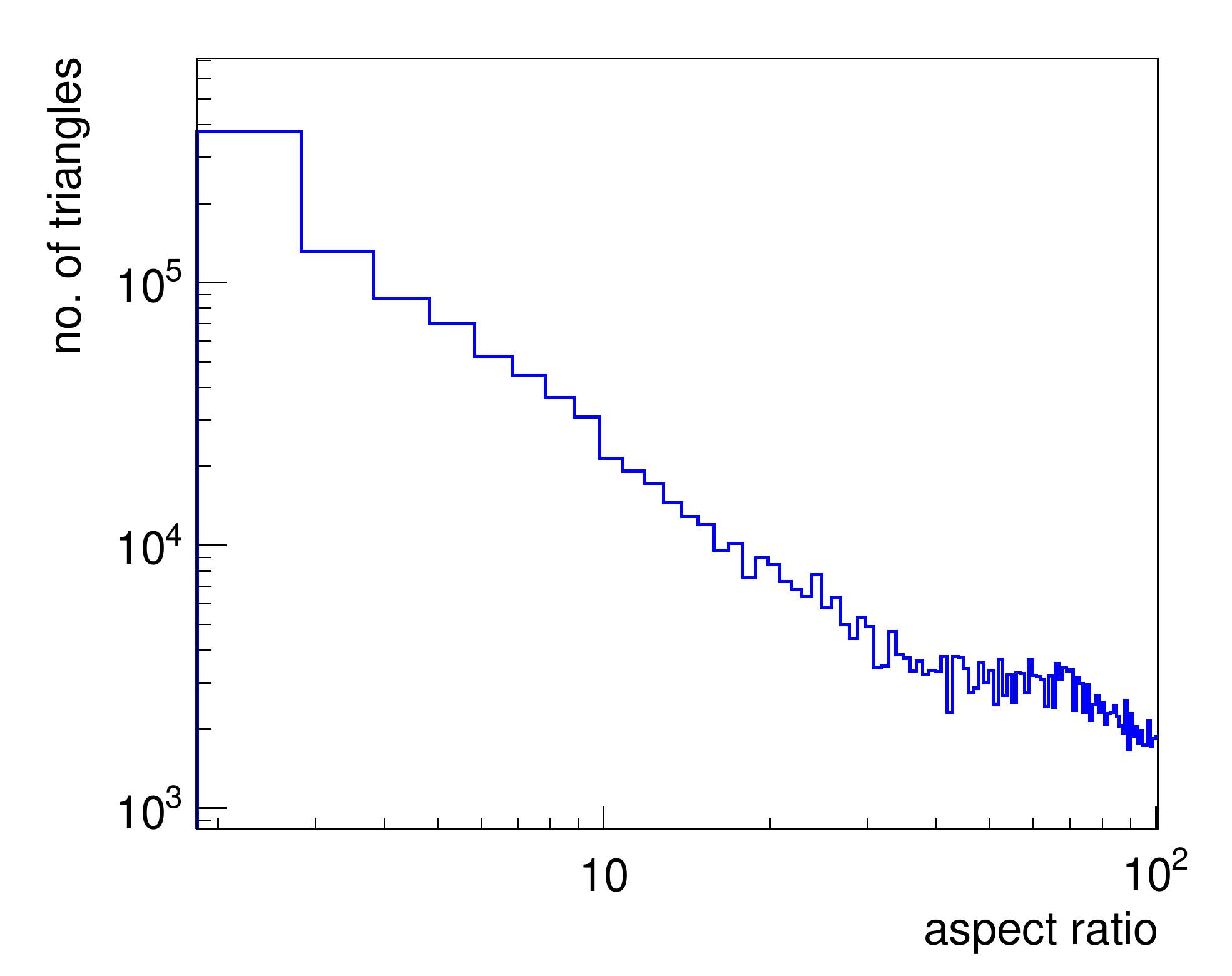}\label{FigARdistTriangles}}\quad
    \subfigure[Rectangles]{\includegraphics[width=0.48\textwidth]{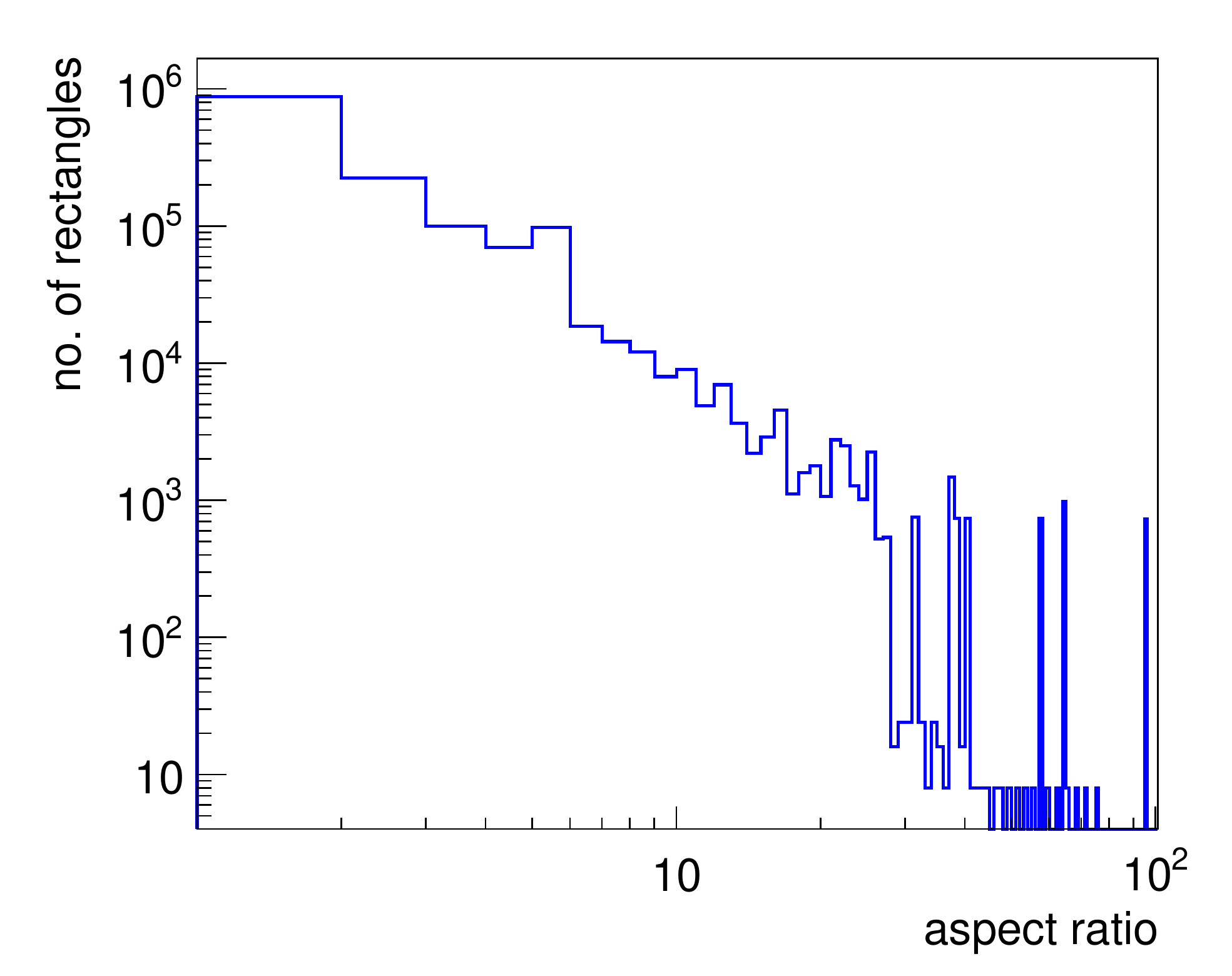}\label{FigARdistRectangles}}         
    \caption{Aspect ratio distribution of triangles (left) and rectangles (right) 
    of our two discretization models of the KATRIN main spectrometer electrode system with electric dipole field.
    Vertical axis: number of elements per bin, with constant bin size 1.  }
    \label{FigARdist}
\end{figure}

\begin{figure}[htbp]
    \centering
    \subfigure[Triangles]{\includegraphics[width=0.48\textwidth]{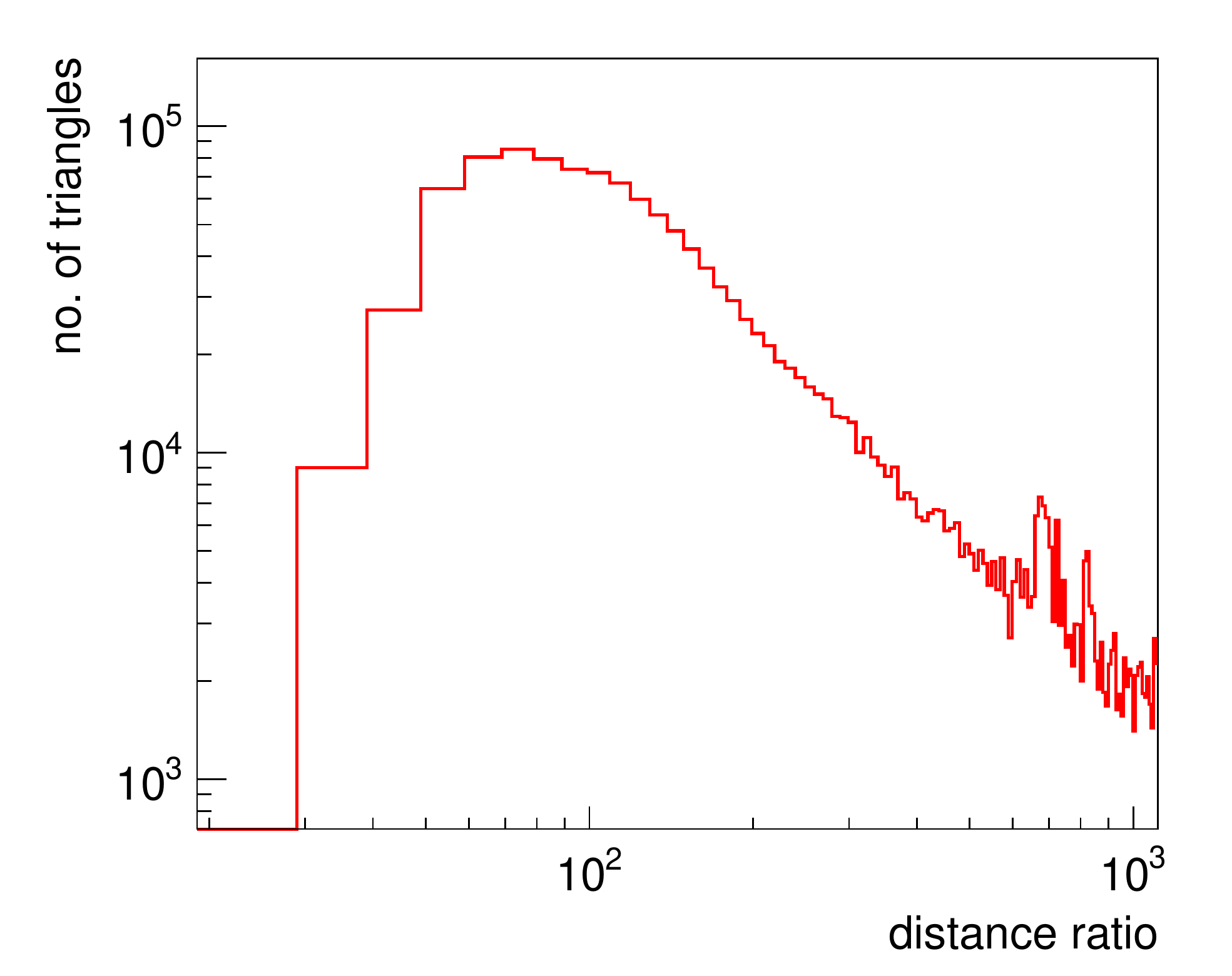}\label{FigDRdistTriangles}}\quad
    \subfigure[Rectangles]{\includegraphics[width=0.48\textwidth]{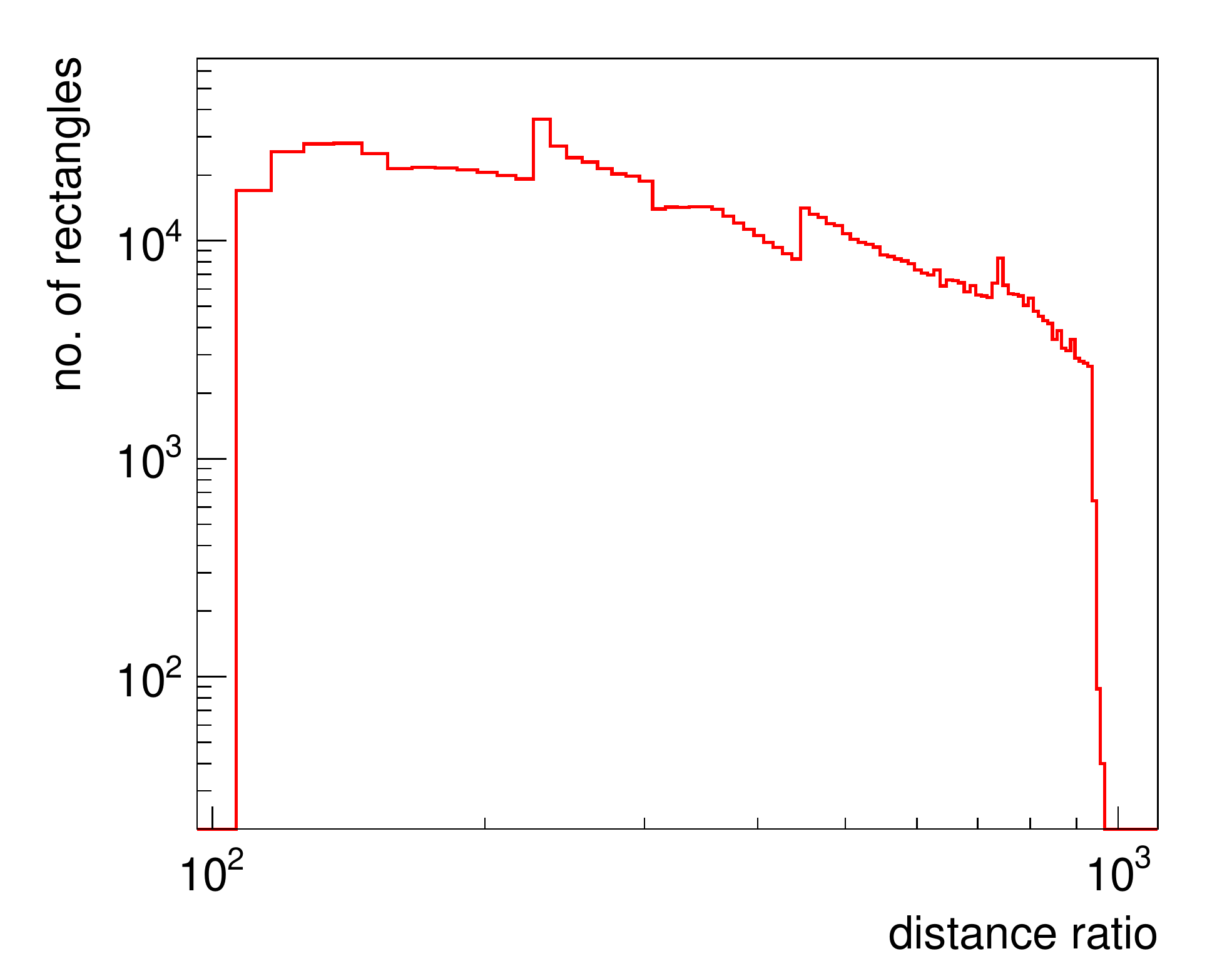}\label{FigDRdistRectangles}}         
    \caption{Distance ratio distribution of triangles (left) and rectangles (right) 
    of our two discretization models of the KATRIN main spectrometer electrode system with electric dipole field,
    at field point ${\bf P}=(0,0,0)$ (center of spectrometer).
    Vertical axis: number of elements per bin, with constant bin size 10. }
    \label{FigDRdist}
\end{figure}

Figure \ref{FigARdist} presents the aspect ratio distribution of the two models 
(left: triangles, right: rectangles). One can see that especially the triangle model has many triangles with large
(${\rm AR}>20$) aspect ratios.
Figure \ref{FigDRdist} shows the distance ratio distributions for these two models for the central field point
${\bf P}=(0,0,0)$. Due to the large number of elements and the small element sizes (relative to typical
field point -- element distances), most of the distance ratio values  are above 100, where the fast
7-point cubature method can be applied.

The electrode discretization 
has been assembled with the software tool \textit{KGeoBag}, which is a C++ library 
allowing to define arbitrary three-dimensional electrode geometries in a text file based on the Extensible 
Markup Language (XML) \cite{TJdiss,Furse,Groh,Combe}. The meshed geometry data, which contains also information 
on the applied  electric potentials, can be automatically created from the input and transferred to the field
computation program \textit{KEMField}. The latter \cite{TJdiss,TJmaster}
is a library, written in C++, which can solve electrostatic and magnetostatic problems with  
BEM. The electrostatics code part contains several logically separated classes which are responsible for

\begin{itemize}
    \item[a)] computation of boundary integrals over single surface (mesh) elements;
    \item[b)] linear algebra routines for determining the charge densities by solving a linear algebraic equation system.
    \item[c)] Integrating field solver, and
    \item[d)] field solving with fast Fourier transform on multipoles (FFTM) \cite{BarrettPhD,Gosda,Barrett}.
\end{itemize}

\textit{KEMField} profits from a very small memory footprint through usage of shell matrices that are not stored in 
memory, allowing to solve for
arbitrary large equation systems with the Robin Hood iterative method 
\cite{Formaggio}.
Furthermore KEMField profits from libraries and computing languages, like MPI
 (Messaging Parsing Interface) \cite{MPI} and OpenCL \cite{Stone} in order to gain computation speed from parallel 
platforms like multi-core CPUs and highly parallel graphical processors.

With the computed charge density values, the electric potential and field at an arbitrary field point can be
calculated by summing the potential and field contributions of the individual elements.
In order to compare the relative errors of the various field computation methods, we generated randomly
3000 field points inside a cylinder with 9 m length and 3.5 m radius (centered at the KATRIN 
main spectrometer vessel center), and we computed the average of the relative errors defined in Secs.
\ref{SectionAnalytical} and \ref{SectionPotentialField}.
As reference potential and field values, we used the Gauss-Legendre biquadrature method desribed in
Sec. \ref{SectionAnalytical}.
Table \ref{TableErrorComparison} summarizes our results for two different analytical methods and for Gaussian
cubature. In the case of triangles, we performed two calculations: first, with all triangles, and second, using only the
small aspect ratio (${\rm AR}<10$) triangles (since we showed in Secs. \ref{SectionAnalytical} and 
\ref{SectionPotentialField} that both the analytical and the Gaussian cubature calculations for triangles are sensitive to
the aspect ratio). One can see in table  \ref{TableErrorComparison}  the following features:  a, the Gaussian cubature
 integration method has much smaller relative errors than both analytical integrations: the potential error is 
practically double precision, while the field error is somewhat larger, but also close to the double precision level;
b, the analytical method 2 (Refs. \cite{Hanninen,Hilkdiss}) has smaller errors than the analytical method 1
 (Refs. \cite{TJdiss,Formaggio});
c, the potential errors are in all cases smaller than the field errors; d, in case of using only smaller aspect ratio triangles,
the errors are smaller than with all (i.e. also large aspect ratio) triangles.
We mention here that we tried to reduce the potential and field rounding errors of the 1.5 million elements
by Kahan summation \cite{Ueberhuber,Kahan}, but with no success.

In addition to the large relative potential and field errors, 
the analytical 1 method of Refs. \cite{TJdiss,Formaggio} has the
following problem for the field  computation of triangles: 
at some very sharply defined field points (e.g. ${\bf P}=(0,0,0)$)
the relative field error is extremely large (more than 10), i.e. the field value
is completely wrong, and in some cases one gets nan or inf results (with C++ compiler). Only a few triangles
(from the 1.5 million) are responsible for these wrong field values (they have small -- ${\rm AR}<10$ -- 
aspect ratio).

\begin{table}[!htbp]
    \begin{center}
    \caption{Average relative error of potential and field simulation of triangles and rectangles at 3000 field points,
    computed with two different analytical methods and with Gaussian cubature.
    Analytical 1: Refs. \cite{TJdiss,Formaggio};  Analytical 2: Refs. \cite{Hanninen,Hilkdiss}. 
    \label{TableErrorComparison}} 
    \vspace{2mm}
        \begin{tabular}{lccc}
        \toprule
                                                    &Analytical 1               &  Analytical 2                  & Gaussian cubature \\
        \midrule
        Potential error (triangles)       &  $7.1\cdot 10^{-9}$   &  $4.6\cdot 10^{-12}$  &  $4.9\cdot 10^{-16}$   \\ 
        Field error  (triangles)           & $1.8\cdot 10^{-6}$     &  $3.2\cdot 10^{-9}$    & $4.5\cdot 10^{-14}$                   \\
        \midrule
        Potential error (triangles, ${\rm AR}<10$)   & $1.8\cdot 10^{-11}$  & $2.0\cdot 10^{-12}$  & $3.2\cdot 10^{-16}$ \\ 
        Field error  (triangles, ${\rm AR}<10$)        & $1.4\cdot 10^{-7}$   & $6.7\cdot 10^{-10}$   & $2.4\cdot 10^{-14}$                    \\ 
        \midrule
        Potential error (rectangles)       & $2.6\cdot 10^{-11}$  & $1.7\cdot 10^{-11}$  &  $1.1\cdot 10^{-16}$       \\ 
        Field error  (rectangles)           & $6.3\cdot 10^{-7}$   &  $5.2\cdot 10^{-9}$    &  $1.7\cdot 10^{-14}$     \\ 
        \bottomrule
        \end{tabular}
    \end{center}
\end{table}

\section{Computation time with CPU and GPU}
 \label{SectionCompTime}

In this section, we compare the computer speed of the Gaussian cubature and the analytical integration 
methods on CPU (C++) and GPU (OpenCL) \cite{Stone, Hwu}.
For this purpose, we calculated the electric potential and 
field of the 2 electrode models described in the previous section and containing
1.5 million triangles and 1.5 million rectangles, respectively.
Table \ref{TableCompTimeCPU} presents the CPU computation time values
for 100 field points and five different calculation types: two analytical 
(Refs. \cite{Hanninen,Hilkdiss,TJdiss,Formaggio}) and three Gaussian cubature
methods. At the cubature method, the Gaussian points are calculated from
 the individual element geometry before each potential / field calculation.
One can see that the Gaussian cubature methods (especially those with 7 and 12 points)
are significantly faster than the analytical calculations. 
The computation time of the Gaussian cubature formulas increases almost linearly
with the number of Gaussian points.
The triangle and the rectangle integrations have
approximately the same speed.

\begin{table}[!htbp]
    \begin{center}
    \caption{Computation time values and speed increase factors (relative to analytical speed)
    for cubature implementation on CPU, with field and
    potential of 1.5 million triangles and rectangles
    computed at 100 field points.
    Analytical 1: Refs. \cite{TJdiss,Formaggio};  Analytical 2: Refs. \cite{Hanninen,Hilkdiss}. 
    SF1: (time of analytical 1)/time; SF2: (time of analytical 2)/time.
    \label{TableCompTimeCPU}} 
    \vspace{2mm}
        \begin{tabular}{lllll}
        \toprule
        Element type & Computation method  & Time (s) & SF1 & SF2   \\
        \midrule
        Triangles   &  Analytical 1 &   161    & 1 & 0.44 \\ 
            &  Analytical 2   &   70 & 2.3  & 1   \\ 
            &  7-point cubature   &  15.7 & 10.3 & 4.5     \\ 
            &  12-point cubature   &  25.5 & 6.3  & 2.7    \\ 
            &  33-point cubature   &  61 & 2.6 &  1.1    \\
        \midrule
        Rectangles   &  Analytical 1 &   140    & 1 & 0.56   \\ 
            &  Analytical 2   &   79 & 1.8  &  1  \\ 
            &  7-point cubature   &  18 & 7.7 & 4.4    \\ 
            &  12-point cubature   & 28  &  5.1  &  2.8  \\ 
            &  33-point cubature   & 70  &  2.0  & 1.1   \\
        \bottomrule
        \hline
        \end{tabular}
    \end{center}
\end{table}

Table \ref{TableCompTimeGPU} shows the time comparisons  on a
 GPU with OpenCL,  with two different
 analytical methods (Refs. \cite{Hanninen,Hilkdiss,TJdiss,Formaggio})
 and with a distance ratio dependent cubature integrator incorporating the 
7-point, 12-point  and 33-point cubature methods.
Also on this platform the cubature is much faster than analytical methods 
and can deliver a speed up of almost an order of magnitude for triangles.

\begin{table}[!htbp]
    \begin{center}
    \caption{Computation time values and speed factors for cubature implementation on GPU, with field and
    potential computed at 10000 field points.
    Analytical 1: Refs. \cite{TJdiss,Formaggio};  Analytical 2: Refs. \cite{Hanninen,Hilkdiss}. 
    In the case of distance ratio dependent computation
    the 7-point, 12-point and the 33-point cubature methods and the analytical method 2 are used.
    \label{TableCompTimeGPU}} 
    \vspace{2mm}
        \begin{tabular}{llll}
        \bottomrule
        Element type & Computation method  & Time (s) & Speed increase factor   \\  
                                &  &  & (relative to analytical 1)  \\
        \midrule
        Rectangles   &  Analytical 1 &   82.7    & 1  \\ 
            &  Analytical 2   &   81.7 & 1.01     \\ 
            &  7-point cubature   &  27.8 & 3     \\ 
            &  Distance ratio dependent   & 64.8  &  1.3     \\
        \midrule
        Triangles   &  Analytical 1 &   258    & 1  \\ 
            &  Analytical 2   &   80.2 & 3.2     \\ 
            &  7-point cubature   &  28.3 & 9.1     \\ 
            &  Distance ratio dependent   & 97  &  2.7     \\ 
        \bottomrule
        \end{tabular}
    \end{center}
\end{table}

We investigated also a possible time benefit by saving the Gaussian points for all elements into heap 
memory in advance. The storage of the Gaussian points can require a lot of memory (e.g.: 800 MB for 5
 million element and 7 points for each element), depending from the meshed input geometry.
In the following we compare the 7-point and 12-point cubature against the analytical method as discussed
 in \cite{Hanninen}. 
As shown in table \ref{TableCompTimeCahed}, 
again we gather a speed increase up to factor five by using the Gaussian cubature 
versus the analytical 2 method of \cite{Hanninen}. Since the Gaussian points are computed on the fly in the 
fast stack memory, the non-cached variant of our code is only marginally slower than the code version
 with precomputed Gaussian points.

\begin{table}[!htbp]
    \begin{center}
    \caption{Speed test with 1.5 Million triangles, 100 field points, precomputed vs. non-precomputed Gaussian points.
    \label{TableCompTimeCahed}} 
    \vspace{2mm}
        \begin{tabular}{lll}
        \bottomrule
        Computation method  & Time (s) & Speed increase factor   \\  
                                &  & (relative to analytical 2)  \\
        \midrule
        Analytical 2 &   60.7    & 1  \\ 
        7-point cubature, precomputed   &  12.7 & 4.8     \\ 
        7-point cubature, non-precomputed   &  14.5 & 4.2     \\ 
        12-point cubature, precomputed   &  19.9 & 3     \\ 
        12-point cubature, non-precomputed   &  23 & 2.6     \\ 
        \bottomrule
        \end{tabular}
    \end{center}
\end{table}

We recognized that the Gaussian cubature computation time depends very much on implementation details.
In the C++ codes of the above described calculations we used double arrays for the representation of
the Gaussian point, field point and electric field components. In another calculation, we used the TVector3
class  of the \textit{ROOT} data analysis code package \cite{ROOT}. The code layout is more elegant and clean by using the
TVector3 class, but in this case  the  computation is four times slower than by using double arrays.
Interestingly, the cubature computation is about two times slower even if the TVector3 commands are inside the C++
functions but they are not used for the calculations.

GPU architectures profit from a high degree of parallelism, even though the clock speed is not that high as on
 CPUs. In order to guarantee a highly parallel execution of the code, the used data fragments may not
 be too large, hence we have to avoid large double arrays, because using too large data arrays results 
in a lower degree of parallelism (and hence less speed) (e.g. for the 33-point cubature the array containing
 the Gaussian points is 99-dimensional). This is why we focused on
 avoiding saving data in large arrays due to limitation of register memory on GPU chips. Instead, we are using mostly single double values achieving a very
 highly parallel execution of the code.\\

At the end of this section, we present a few technical details about the computers and codes that we used.
We run all our single-threaded C++ code on an Intel Xeon CPU with 3.10 GHz clock speed (E5-2687W v3).
 All CPU programs have been compiled with GCC 4.8.4 with optimization flag O3.
Our code has been ported to OpenCL (version 1.1) as well. In order to test the speed on 
GPUs, we use a Tesla K40c card running at 875 MHz.

\section{Conclusions}
 \label{SectionConclusions}

Integration over triangles and rectangles is important for many applications of mathematics, science 
and engineering, especially for FEM and BEM. 
Analytical integration is believed to be more accurate than numerical integration.
Nevertheles, in some special cases the latter one has higher accuracy and also higher speed.
We demonstrated in our paper that in the case of electric potential and field calculation of charged triangles
and rectangles at points far from these elements the Gaussian cubature numerical integration method
is much more accurate and faster than some of the best analytical integration methods. Using the Gaussian
cubature method, the triangles and rectangles with continuous charge distribution are replaced
by discrete point charges, the potential and field of which can be computed by simple formulas.
At field points far from the elements, the analytical methods have large rounding errors, while the accuracy of
the Gaussian cubature method is limited only by computer arithmetic precision. Closer to the elements, a
Gaussian cubature formula with higher number of Gaussian points (nodes) has to be employed, in order to get the
maximal accuracy. Very close to the elements, the Gaussian cubature method is not precise enough, 
therefore analytical integration has to be used there. Nevertheless, for a typical BEM problem the field point
is far from most of the elements, therefore the simple, fast and accurate Gaussian cubature method can be used
for a large majority of the boundary elements.

The complex electrode examples described in Sec. \ref{SectionManyElements} illustrate that the Gaussian 
cubature method can be four to seven orders of magnitude more accurate than some of the best analytical methods
that can be found in the literature.
In addition, the examples in Sec. \ref{SectionCompTime} show that the potential and field computation with
Gaussian cubature can be three to ten times faster than the analytical methods, both with CPU and with GPU.
An additional advantage of the Gaussian cubature method is that accurate higher derivatives of the electric field
can be relatively easily calculated analytically by point charges, while in the case of analytical integrations
this is a rather difficult task. The higher derivatives can be useful for field mapping computations
in conjunction with the Hermite interpolation method.

In our paper, we compared the Gaussian cubature method with analytical integration in the case of constant
BEM elements (i.e. elements with constant charge density). Nevertheless, the Gaussian cubature method can
be easily applied also for elements with arbitrary charge density function (e.g. linear, quadratic etc.). 
Most probably,
the Gaussian cubature method is more accurate and faster than analytical integration also in the case of
higher order charge density functions. 
We conjecture that the Gaussian cubature method presented in our paper for electrostatics
can also be applied for magnetostatics and time-dependent electromagnetic problems.
E.g. magnetic materials can be computed by fictive magnetic charges,
and the magnetic field in that case can be calculated by similar formulas than electric field of
electric charges.
It might be that the Gaussian cubature numerical integration method
can also be used for the efficient computation of multipole moment  coefficients of triangles and rectangles
(see Ref. \cite{Barrett} for analytical integration results).
The ten different Gaussian cubature formulas presented in our paper can be used for arbitrary high-precision 
and fast integrations over triangles and rectangles.

\subsection*{Acknowledgments}

We are grateful to Prof. J. Formaggio, Dr. T. J. Corona and J. Barrett for many fruitful discussions,
 especially on the comparisons of the numerical results against the analytical method of Refs. \cite{Hanninen,Hilkdiss},
and we thank Prof. J. Formaggio and J. Barrett for their reading the manuscript and important comments.
Furthermore, we thank Prof. G. Drexlin for his continuous support of this dedicated project.
D. Hilk would like to thank the Karlsruhe House Of Young Scientists (KHYS) for a research travel grant to
 the Laboratory of Nuclear Science at MIT in order to work on the cubature implementation on graphical
 processors together with Prof. Joseph Formaggio and John Barrett.
We acknowledge the support of the German Helmholtz Association HGF and the German Ministry for 
Education and Research BMBF (05A14VK2 and 05A14PMA).

\begin{appendices}

\newpage

\section{Gaussian points and weights for triangles}
\label{AppendixTriangles}

\setcounter{table}{0}
\renewcommand{\thetable}{A\arabic{table}}

\begin{table}[!htbp]
    \begin{center}
    \caption{Barycentric coordinates and weights of the  4-point (degree 3) Gaussian cubature for triangle.
    See Refs. 
    \cite{Stroud,HammerStroud1956,HammerStroud1958}.
    \label{TableTriangle4}} 
    \vspace{2mm}
        \begin{tabular}{cccc}
        \toprule
        $\lambda_A$   & $\lambda_B$  & $\lambda_C$  & weight \\
        \midrule
            1/3   &    1/3                &    1/3       &    -9/16       \\ 
            3/5  &    1/5       &  1/5      &   25/48   \\ 
            1/5  &    3/5       &  1/5      &   25/48   \\ 
            1/5  &    1/5       &  3/5      &   25/48   \\ 
        \bottomrule
        \end{tabular}
    \end{center}
\end{table}

\begin{table}[!htbp]
    \begin{center}
    \caption{Barycentric coordinates and weights of the 7-point (degree 5) Gaussian cubature for triangle;
    $t=1/3$, $s=(1-\sqrt{15})/7$, $r=(1+\sqrt{15})/7$.
    Both the  second  and the third row corresponds to three Gaussian points with equal weights, according to all possible
    different permutations of $\lambda_A$ and $\lambda_B=\lambda_C$ (like the rows two to four in Table \ref{TableTriangle4}).
    From Refs. 
    \cite{Radon,HammerMarloweStroud,Stroud},  \cite{EngelnMullges} (p. 420).
    \label{TableTriangle7}} 
    \vspace{2mm}
        \begin{tabular}{cccc}
        \toprule
        $\lambda_A$   & $\lambda_B$  & $\lambda_C$  & weight \\
        \midrule
            $t$   &    $t$                &    $t$       &    9/40       \\ 
            $t+2ts$  &    $t-ts$       &  $t-ts$      &   $(155+\sqrt{15})/1200$    \\ 
            $t+2tr$    &  $t-tr$         &   $t-tr$       &   $(155-\sqrt{15})/1200$    \\ 
        \bottomrule
        \end{tabular}
    \end{center}
\end{table}

\begin{table}[!htbp]
    \begin{center}
    \caption{Barycentric coordinates and weights of the 12-point (degree 7) Gaussian cubature for triangle.
    Each row corresponds to three Gaussian points with equal weights: we get the second and third points
    by the permutations $(\lambda_A,\lambda_B,\lambda_C) \to  (\lambda_B,\lambda_C,\lambda_A)$
    and $(\lambda_A,\lambda_B,\lambda_C) \to  (\lambda_C,\lambda_A,\lambda_B)$, respectively;
    taken from Ref. \cite{Gatermann}.
    \label{TableTriangle12}} 
    \vspace{2mm}
        \begin{tabular}{cccc}
        \toprule
        $\lambda_A$   & $\lambda_B$  & $\lambda_C$  & weight \\
        \midrule
        0.06238226509439084     &   0.06751786707392436   &   0.8700998678316848  &  0.05303405631486900         \\ 
        0.05522545665692000     &   0.3215024938520156    &   0.6232720494910644     & 0.08776281742889622         \\ 
        0.03432430294509488      &   0.6609491961867980    &  0.3047265008681072      &  0.05755008556995056      \\ 
            0.5158423343536001          &    0.2777161669764050    &   0.2064414986699949    &  0.13498637401961758        \\ 
        \bottomrule
        \end{tabular}
    \end{center}
\end{table}

\begin{table}[!htbp]
    \begin{center}
    \caption{Barycentric coordinates and weights of the  19-point (degree 9) Gaussian cubature for triangle.
    Each of the  rows two to five corresponds to three Gaussian points with equal weights, according to all possible
    different permutations of $\lambda_A$ and $\lambda_B=\lambda_C$ (like in Table \ref{TableTriangle4}).
    The last row corresponds to six Gaussian points with equal weights, according to all possible
    permutations of $\lambda_A$, $\lambda_B$ and $\lambda_C$. 
    Numbers taken from Ref. \cite{LynessJespersen}.
    \label{TableTriangle19}} 
    \vspace{2mm}
        \begin{tabular}{cccc}
        \toprule
        $\lambda_A$   & $\lambda_B$  & $\lambda_C$  & weight \\
        \midrule
        1/3   & 1/3     &   1/3     &  0.09713579628279610         \\ 
        0.02063496160252593      &  0.48968251919873704     &  0.48968251919873704      &   0.03133470022713983       \\ 
        0.1258208170141290     &  0.4370895914929355     &  0.4370895914929355      &    0.07782754100477543    \\ 
        0.6235929287619356     &  0.18820353561903219      &  0.18820353561903219     &    0.07964773892720910      \\ 
        0.9105409732110941     &  0.04472951339445297     &   0.04472951339445297     &    0.02557767565869810    \\ 
        0.03683841205473626    &  0.7411985987844980      &   0.22196298916076573    &    0.04328353937728940      \\ 
        \bottomrule
        \end{tabular}
    \end{center}
\end{table}

\begin{table}[!htbp]
    \begin{center}
    \caption{Barycentric coordinates and weights of the 33-point (degree 12) Gaussian cubature for triangle.
    Each of the first five rows corresponds to three Gaussian points with equal weights, according to all possible
    different permutations of $\lambda_A$ and $\lambda_B=\lambda_C$ (like in Table \ref{TableTriangle4}).
    Each of the last three rows corresponds to 6 Gaussian points with equal weights, according to all possible
    permutations of $\lambda_A$, $\lambda_B$ and $\lambda_C$. 
    Numbers taken from Ref. \cite{Papanicolopulos}.
    \label{TableTriangle33}} 
    \vspace{2mm}
        \begin{tabular}{cccc}
        \toprule
        $\lambda_A$   & $\lambda_B$  & $\lambda_C$  & weight \\
        \midrule
        0.4570749859701478   &  0.27146250701492611   &  0.27146250701492611  & 0.06254121319590276   \\
        0.1197767026828138   & 0.44011164865859310   &  0.44011164865859310  & 0.04991833492806094   \\
        0.0235924981089169  & 0.48820375094554155   &  0.48820375094554155  & 0.02426683808145203   \\
        0.7814843446812914   & 0.10925782765935432   &  0.10925782765935432  & 0.02848605206887754   \\
        0.9507072731273288  & 0.02464636343633558   &  0.02464636343633558  & 0.00793164250997364   \\
        0.1162960196779266   & 0.2554542286385173   &  0.62824975168355610  & 0.04322736365941421   \\
        0.02303415635526714   & 0.2916556797383410   &  0.68531016390639186  & 0.02178358503860756   \\
        0.02138249025617059   &   0.1272797172335894   &  0.85133779251024000  & 0.01508367757651144   \\
        \bottomrule
        \end{tabular}
    \end{center}
\end{table}

\newpage

\section{Gaussian points and weights for rectangles}
\label{AppendixRectangles}

\setcounter{table}{0}
\renewcommand{\thetable}{B\arabic{table}}

\begin{table}[!htbp]
    \begin{center}
    \caption{Natural coordinates and weights of the 4-point (degree 3) Gaussian cubature for rectangle;
    $s=1/\sqrt{3}$.
    From Refs. 
    \cite{EngelnMullges} (p. 417), \cite{AlbrechtCollatz} (p. 8).
    \label{TableRectangle4}} 
    \vspace{2mm}
        \begin{tabular}{cccc}
        \toprule
        $x$   & $y$  &  weight \\
        \midrule
            s   &    s       &    1/4       \\ 
            s   &   -s       &  1/4          \\ 
            -s   &   s       &    1/4       \\ 
            -s   &   -s       &  1/4          \\ 
        \bottomrule
        \end{tabular}
    \end{center}
\end{table}

\begin{table}[!htbp]
    \begin{center}
    \caption{Natural coordinates and weights of the  7-point (degree 5) Gaussian cubature for rectangle;
    $t=\sqrt{14/15}$, $r=\sqrt{3/5}$,  $s=\sqrt{1/3}$.
    The last row corresponds to 4 Gaussian points with equal weights, according to all possible
    sign changes (like  in Table \ref{TableRectangle4}): $(x,y)\to \; (r,s),\, (r,-s),\, (-r,s),\, (-r,-s)$.
    From Refs. 
    \cite{Stroud} (p. 246), \cite{Radon} (p. 298), \cite{AlbrechtCollatz} (p. 9).
    \label{TableRectangle7}} 
    \vspace{2mm}
        \begin{tabular}{cccc}
        \toprule
        $x$   & $y$  &  weight \\
        \midrule
            0   &    0       &    2/7       \\ 
            0   &   t       &  5/63          \\ 
            0   &   -t       &    5/63       \\ 
            r   &   s       &  5/36          \\ 
        \bottomrule
        \end{tabular}
    \end{center}
\end{table}

\begin{table}[!htbp]
    \begin{center}
    \caption{Natural coordinates and weights of the 12-point (degree 7) Gaussian cubature for rectangle;
    $r=\sqrt{6/7}$, $s=\sqrt{(114-3\sqrt{583})/287}$,  $t=\sqrt{(114+3\sqrt{583})/287}$,
    $B_1=49/810$, $B_2=(178981+2769\sqrt{583})/1888920$, $B_3=(178981-2769\sqrt{583})/1888920$.
    Each of the last two rows corresponds to four Gaussian points with equal weights, according to all possible
    sign changes of $s$ and $t$, like  in Tables \ref{TableRectangle4} and \ref{TableRectangle7}.
    From Refs. \cite{Stroud} (p. 253), \cite{Tyler} (p. 403).
    \label{TableRectangle12}} 
    \vspace{2mm}
        \begin{tabular}{cccc}
        \toprule
        $x$   &   $y$      &  weight \\
        \midrule
            r   &    0       &    $B_1$       \\ 
            -r   &   0       &   $B_1$         \\ 
            0   &   r       &    $B_1$       \\ 
            0   &   -r       &   $B_1$         \\ 
            s   &   s       &   $B_2$         \\ 
            t   &   t       &   $B_3$         \\ 
        \bottomrule
        \end{tabular}
    \end{center}
\end{table}

\begin{table}[!htbp]
    \begin{center}
    \caption{Natural coordinates and weights of the  17-point (degree 9) Gaussian cubature for rectangle.
    Each of the last four rows corresponds to four Gaussian points with equal weights, according to 
    the transformations  $(x,y) \to \; \pm (x,y), \; \pm(-y,x)$ (rotations by $0^\circ$, $180^\circ$  
    and  $\pm 90^\circ$ in the $x-y$ plane).
    From Refs. \cite{Moller} (p. 194), \cite{Engels} (p. 257),  \cite{EngelnMullges} (p. 419).
    \label{TableRectangle17}} 
    \vspace{2mm}
        \begin{tabular}{cccc}
        \toprule
        $x$   &   $y$      &  weight \\
        \midrule
        0                                      &  0                                          &      0.131687242798353921        \\ 
        0.968849966361977720    &  0.630680119731668854        &      0.022219844542549678      \\ 
        0.750277099978900533    &  0.927961645959569667        &     0.028024900532399120     \\ 
        0.523735820214429336    &  0.453339821135647190        &     0.099570609815517519       \\ 
        0.076208328192617173    &  0.852615729333662307        &     0.067262834409945196      \\ 
        \bottomrule
        \end{tabular}
    \end{center}
\end{table}

\clearpage
\begin{table}[!htbp]
    \begin{center}
    \caption{Natural coordinates and weights of the  33-point (degree 13) Gaussian cubature for rectangle.
    Each of the last eight rows corresponds to four Gaussian points with equal weights, according to 
    the transformations  $(x,y) \to \; \pm (x,y), \; \pm(-y,x)$ (rotations by $0^\circ$, $180^\circ$  
    and  $\pm 90^\circ$ in the $x-y$ plane).
    From Ref. \cite{CoolsHaegemans} (p. 145).
    \label{TableRectangle33}} 
    \vspace{2mm}
        \begin{tabular}{cccc}
        \toprule
        $x$   &   $y$      &  weight \\
        \midrule
        0                                   &  0                                    &   0.075095528857806335   \\ 
        0.778809711554419422     &  0.983486682439872263        &    0.007497959716124783        \\ 
        0.957297699786307365     &  0.859556005641638928        &    0.009543605329270918        \\ 
        0.138183459862465353     &  0.958925170287534857        &    0.015106230954437494        \\ 
        0.941327225872925236     &  0.390736216129461000        &    0.019373184633276336        \\ 
        0.475808625218275905     &  0.850076673699748575        &    0.029711166825148901        \\ 
        0.755805356572081436     &  0.647821637187010732        &    0.032440887592500675       \\ 
        0.696250078491749413     &  0.070741508996444936        &    0.053335395364297350        \\ 
        0.342716556040406789     &  0.409304561694038843        &    0.064217687370491966        \\ 
        \bottomrule
        \end{tabular}
    \end{center}
\end{table}

\end{appendices}

\end{document}